\documentclass[prb,a4paper,aps,twocolumn,eqsecnum,amsmath,amssymb,showpacs]{revtex4}
\usepackage{graphicx,rotating,subfigure,amsmath,amsfonts,amssymb,delarray,pstricks}

\renewcommand{\vec}[1]{\boldsymbol #1}

\def\12{\frac{1}{2}}
\begin{document}
\bibliographystyle{apsrev}

\title{Magnetic excitations in one-dimensional spin-orbital models}

\author{Alexander Herzog}
\affiliation{ Max-Planck-Institut f\"ur Festk\"orperforschung,
              Heisenbergstrasse 1, D-70569 Stuttgart, Germany }

\author{Andrzej M. Ole\'s}
\affiliation{ Max-Planck-Institut f\"ur Festk\"orperforschung,
              Heisenbergstrasse 1, D-70569 Stuttgart, Germany }
\affiliation{ Marian Smoluchowski Institute of Physics, Jagellonian
              University, Reymonta 4, PL-30059 Krak\'ow, Poland }

\author{Peter Horsch}
\affiliation{ Max-Planck-Institut f\"ur Festk\"orperforschung,
              Heisenbergstrasse 1, D-70569 Stuttgart, Germany }

\author{Jesko Sirker} 
\email[]{j.sirker@fkf.mpg.de}
\affiliation{Department of Physics and Research Center
  OPTIMAS, University of Kaiserslautern, D-67663
  Kaiserslautern, Germany} 
\affiliation{Max-Planck-Institut f\"ur Festk\"orperforschung,
  Heisenbergstrasse 1, D-70569 Stuttgart, Germany }

\date{\today}

\begin{abstract}
  We study the dynamics and thermodynamics of one-dimensional
  spin-orbital models relevant for transition metal oxides. 
We show that 
collective spin, orbital, and combined spin-orbital excitations with
infinite lifetime can exist, if the ground state of both sectors is
ferromagnetic.
Our main focus is the case of effectively ferromagnetic
(antiferromagnetic) exchange for the spin (orbital) sector, respectively,
and we investigate the renormalization of spin excitations via spin-orbital
fluctuations using a boson-fermion representation.  We contrast a
mean-field decoupling approach with results obtained by treating the
spin-orbital coupling perturbatively. Within the latter
self-consistent approach we find a significant increase of the
linewidth and additional structures in the dynamical spin structure
factor as well as Kohn anomalies in the spin-wave dispersion caused by
the scattering of spin excitations from orbital fluctuations.  Finally, we
analyze the specific heat $c(T)$ by comparing a numerical solution of
the model obtained by the density-matrix renormalization group with
perturbative results. At low temperatures $T$ we find numerically
$c(T)\sim T$ pointing to a low-energy effective theory with dynamical
critical exponent $z=1$.
\end{abstract}


\maketitle

\section{Introduction}
\label{Intro}
In condensed matter systems the coupling between different degrees of
freedom often plays an important role. The electron-phonon coupling,
for example, can lead to the formation of renormalized quasiparticles,
so-called polarons,\cite{Polaron,Polaron2} as well as to phase
transitions like the Peierls instability.\cite{Peierls} In recent
years, the coupling between fermionic and bosonic degrees of freedom
has also been intensely studied in Bose-Fermi (BF) mixtures of ultracold
quantum
gases.\cite{PhysRevLett.90.170403,PhysRevA.68.023606,PhysRevLett.91.150403,PhysRevLett.93.120404,PhysRevLett.92.050401}

Coupled degrees of freedom also seem to be important in certain
transition metal oxides where the low-lying electronic states (termed
``orbitals'') are not completely quenched so that temperature or
doping can lead to a significant redistribution of the valence
electron density. In insulating materials with partly filled
degenerate orbitals the superexchange between the magnetic degrees of
freedom then becomes a function of the orbital occupation. This leads
to models of coupled spin and orbital degrees of freedom as, for
example, the Kugel-Khomskii model where the orbitals are represented
by a pseudospin.\cite{KugKhom}
LaMnO$_3$,\cite{FeinerOles,TokNag,CMR,KovalevaOles} LaTiO$_3$,
\cite{PhysRevLett.85.3950,PhysRevLett.85.3946,PhysRevLett.91.066403,PhysRevB.68.060401}
and LaVO$_3$ or YVO$_3$,
\cite{TIMR1,TIMR2,PhysRevB.62.R9271,PhysRevLett.86.3879,PhysRevLett.91.257202,PhysRevLett.91.257203,PhysRevB.67.100408,PhysRevLett.100.167205}
are well-known examples for compounds believed to be described by
effective spin-orbital models. They exhibit a wide range of
fascinating effects ranging from colossal magnetoresistance\cite{CMR}
to temperature-induced magnetization reversals.\cite{TIMR1,TIMR2}

Common to all these transistion metal oxides is a lifting of the
fivefold degeneracy of the $d$ orbitals into two
$e_g$ orbitals ($x^2-y^2$ and $3z^2-r^2$) and three $t_{2g}$ orbitals
($xy$, $yz$, and $xz$).  This splitting is due to the perovskite
structure where oxygen ions, O$^{2-}$, form octahedra around the
transition metal ions which are therefore exposed to an approximately
cubic crystal field.  As a consequence, the orbitals pointing towards
the oxygen ions are energetically unfavorable.

In YVO$_3$ the $t_{2g}$ orbitals are occupied by two electrons forming
an effective spin $S=1$ due to large Hund's rule coupling. The
material is an insulator with an interesting phase diagram.
\cite{TIMR1,TIMR2,PhysRevB.62.R9271,PhysRevLett.91.257202} At
temperatures below $77$ K the system is in a $G$-type
antiferromagnetic (AF) phase, i.e., AF in all three directions. In a
range of higher temperatures, $77$ K $<T< 116$ K, the magnetic
structure is $C$-type with spins ordering antiferromagnetically in the
$(a,b)$ plane and ferromagnetically along the $c$ axis.  The
surprising fact that the ferromagnetic (FM) exchange integral in this
phase is much larger than the AF exchange interactions in the $(a,b)$
plane\cite{PhysRevLett.91.257202} was explained by strong orbital
fluctuations along the $c$-axis chains that trigger
ferromagnetism.\cite{PhysRevLett.86.3879} In the $C$-type phase a
neutron scattering study revealed that the magnon dispersion along the
FM $c$-axis chains consists of two branches.  This splitting has been
interpreted as due to a periodic modulation of the FM exchange along
these chains caused by an entropy gain of fluctuating orbital
occupations.\cite{PhysRevLett.91.257202,PhysRevLett.91.257203} Support
for an {\it orbital Peierls\/} effect in this material was given by
numerical
investigations\cite{PhysRevLett.91.257203,PhysRevB.67.100408} and a
mean-field (MF) decoupling approach.\cite{SPPRL}

However, the \textit{dynamics} in such systems cannot easily be
studied numerically and a MF decoupling is unable to explain important
features of coupled spin-orbital degrees of freedom\cite{OlesHorsch} as can be
seen in the following example. Consider the one-dimensional (1D)
spin-orbital
Hamiltonian\cite{PhysRevB.61.6747,PhysRevB.69.104428,PhysRevB.58.10276}
\begin{equation}
\label{FMSU(4)}
  \mathcal{H}=J\sum_j\left(\vec S_j\cdot\vec S_{j+1}+x\right)
\left(\vec\tau_j\cdot\vec\tau_{j+1}+y\right),
\end{equation}
with {\it ferromagnetic} superexchange interaction $J<0$, where $\vec
S_j$ and $\vec\tau_j$ are spin $S$ and pseudospin $\tau$
operators at site $j$, respectively, and $x$ and $y$ are constants.
For general $x$, $y$ the model has an $\text{SU}(2)\otimes \text{SU}(2)$ symmetry and
exhibits an additional $\mathbb{Z}_2$ symmetry, interchanging spin and
orbital sectors, if $x=y$. For $S=\tau=1/2$ and $x=y=1/4$ the symmetry is enlarged to
$\text{SU}(4)$.\cite{LiMa}

\begin{figure}[t!]
	\includegraphics[scale=.7]{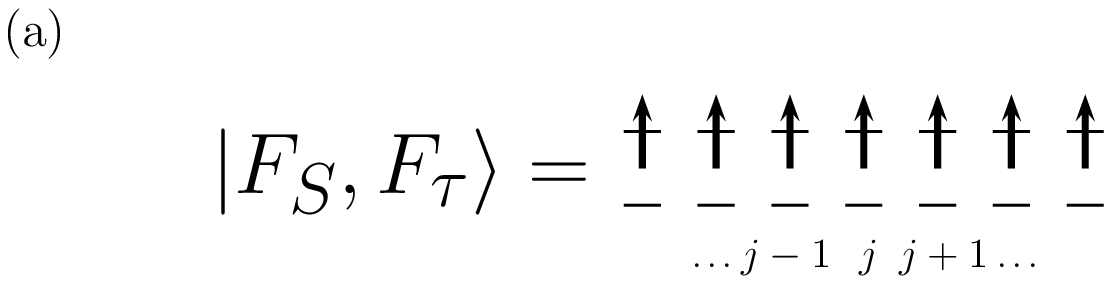}\\*[.7cm]
	\includegraphics[scale=.7]{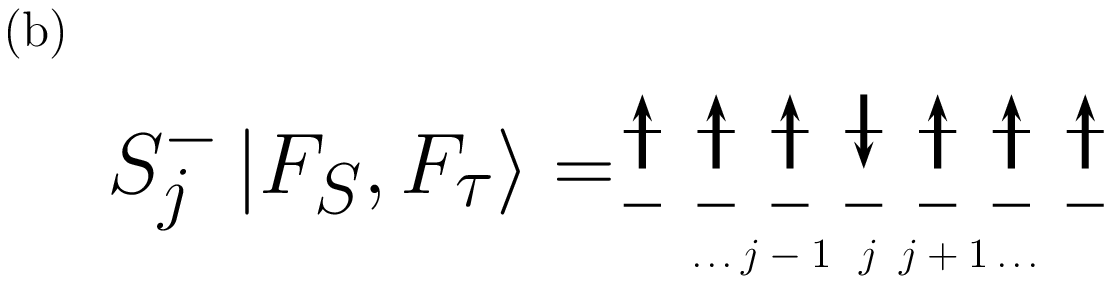}\\*[.7cm]
	\includegraphics[scale=.7]{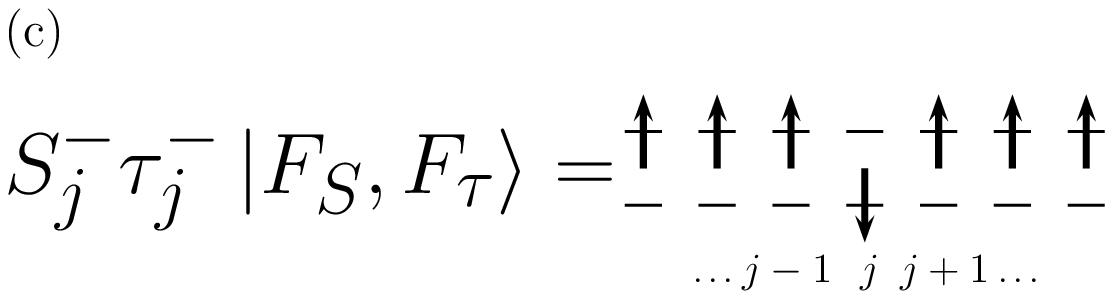}
\caption{(a) Ground state of the FM spin-orbital model,
          Eq.~(\ref{FMSU(4)}), (b) a spin excitation, and (c) a
          coupled spin-orbital excitation. The two orbitals per site
          are assumed to be degenerate (the splitting is only for
          clarity of presentation).}
\label{fig.SU4FM}
\end{figure}

In the following we discuss the case $S=\tau=1/2$ and we choose $x$
and $y$ such that the ground state $\bigl|F_S,F_\tau\bigr>$ is given
by fully polarized spin and orbital sectors as illustrated in
Fig.~\ref{fig.SU4FM}(a).
Using the equation of motion method we find that the state
$S_j^-\bigl|F_s,F_\tau\bigr>$ shown in Fig.~\ref{fig.SU4FM}(b) is
always an elementary excitation with
dispersion $\omega_{S}(q)=|J|(1+4y)(1-\cos q)/4$.
Analogously, the orbital flip is also an elementary excitation with
$\omega_{\tau}(q)=|J|(1+4x)(1-\cos q)/4$. However, these collective excitations are not the
only undamped elementary excitations of the Hamiltonian, Eq.
(\ref{FMSU(4)}). In addition, a coupled spin-orbital
excitation $S_j^-\tau_j^-$, as shown in Fig.~\ref{fig.SU4FM}(c) may exist. To
investigate this issue we again apply the equation of motion method
leading to
\begin{equation}
\left\{\left[\mathcal{H},S_j^-\tau_j^-\right]-|J|\left[C_j(x,y)
+D_j(x,y)\right]\right\}\left|F_S,F_\tau\right>=0,
\end{equation}
where 
$$
C_j(x,y)=(x+y)S_j^-\tau_j^--\frac{1}{4}
\left(S^-_{j-1}\tau^-_{j-1}+S_{j+1}^-\tau_{j+1}^-\right)
$$
is the coherent part, and
\begin{eqnarray}
\label{Dj}
D_j(x,y)&=&-\frac{1}{2}\left[\left(x-\frac{1}{4}\right)
\left(S_j^-\tau_{j-1}^-+S_j^-\tau_{j+1}^-\right)\right. \nonumber \\
&+&\left.\left(y-\frac{1}{4}\right)
\left(S_{j-1}^-\tau_j^-+S_{j+1}^-\tau_j^-\right)\right]
\end{eqnarray}
contains terms which lead to a spatial decoherence of the excitation.
Hence in order to have a fully confined spin-orbital excitation,
$D_j(x,y)$ has to vanish, which obviously is the case if $x=y=1/4$.
From this we draw the conclusion that confined spin-orbital
excitations are rather the exception than the rule, relying on the
particular value of the constants. If we introduce the Bloch states
\begin{equation}
\Psi_{\text{S}\tau}(q)=\frac{1}{N}\sum_j\text{e}^{ijq}S_j^-\tau_j^-\left|F_S,F_\tau\right>,
\end{equation}
the dispersion of the coupled spin-orbital excitation for $x=y=1/4$ is
given by
\begin{equation}
\omega_{\text{S}\tau}(q)=|J|(1-\cos q)/2.
\end{equation}
Thus for $x=y=1/4$ we have
$\omega_S(q)=\omega_\tau(q)=\omega_{\text{S}\tau}(q)$, i.e., the
dispersions of all three elementary excitations are degenerate.\cite{remarkSU4}
Interestingly, they all lie {\it within\/} the continuum of
spin-orbital excitations given by
$\gamma(q,p)=\omega_S(q/2+p)+\omega_\tau(q/2-p)$. The Hamiltonian,
however, does not allow for a decay of these three elementary
excitations in the {\it ferromagnetic} case. For the case of the
coupled spin-orbital excitation we see from Eq.~(\ref{Dj}) that such a
decay becomes possible once we move away from the special point
$x=y=1/4$. Our conclusions partly differ from the ones presented in
Ref.~\onlinecite{PhysRevB.58.10276}, where the coupled spin-orbital
excitation is considered as a bound state below the spin-orbital
continuum.\cite{vdBe}

There are several ways to generalize the $S=\tau=1/2$ case to
arbitrary spin- and pseudospin quantum numbers. If we start again from
fully polarized spin and orbital sectors and only demand that
$S^-_j\tau^-_j\bigl|F_S,F_\tau\bigr>$ stays confined, we find the
condition $x=S(1-S)$ and $y=\tau(1-\tau)$. Another way of generalizing
the $S=\tau=1/2$ case to arbitrary $S$ and $\tau$ relies on the fact
that the Hamiltonian, Eq.~(\ref{FMSU(4)}), with $x=y=1/4$ is equivalent to
\begin{equation}
\label{Dirac}
\mathcal{H}=\frac{J}{4}\sum_j\mathcal{D}_{j,j+1}^{S=\frac{1}{2}}\mathcal{D}_{j,j+1}^{\tau=\frac{1}{2}}.\rm
\end{equation}
where $\mathcal{D}_{j,l}^{\sigma=\frac{1}{2}}$ with $\sigma\in\{S,\tau\}$ is Dirac's exchange operator for $\sigma=1/2$.\cite{Dirac} A generalization
of this exchange operator to arbitrary spin has been discussed by
Schr\"odinger.\cite{Schroedinger} For instance for $\sigma=1$ the spin (pseudospin) exchange operator is given by
$\mathcal{D}^{\sigma=1}_{j,l}=(\vec{\sigma}_j\cdot\vec{\sigma}_l)^2
+\vec\sigma_j\cdot\vec\sigma_l-1$.\cite{PhysRevB.31.3118} The
Hamiltonian in Eq.~(\ref{Dirac}) with arbitrary spin (pseudospin) quantum number does not only keep the single
spin-orbital flip confined as in the generalization discussed above
but rather all spin-orbital excitations of the type
$(S_j^-)^{m_S}(\tau_j^-)^{m_\tau}\bigl|F_S,F_\tau\bigr>$ where $m_S$
and $m_\tau$ are the multiplicities for spin and pseudospin,
respectively. Since a MF decoupling solution treats the spin-orbital
chain as two separate chains with effective exchange parameters
determined self-consistently, the physics of coupled spin-orbital
excitations cannot be captured within this approach.

The purpose of this paper is to study the importance of coupled
spin-orbital excitations in a spin-orbital model with
\textit{antiferromagnetic} superexchange and anisotropic orbital
exchange.  This case is intriguing as spin and orbital degrees of
freedom may be expected to be strongly entangled.\cite{OlesHorsch} In
fact, it has been shown that composite spin-orbital excitations have
to be analyzed together with spin waves in systems with active $e_g$
orbitals, such as for instance KCuF$_3$.\cite{Fei98,Ole00} This
follows from the non-conservation of the orbital flavor in hopping
processes which implies that spin excitations are not independent and
may occur in general together with an orbital flip.
Here we will consider an anisotropic generalization of the
spin-orbital model (\ref{FMSU(4)}) with parameters $x$, $y$ such that
the spins still order ferromagnetically in the ground state. The
orbital sector, however, will no longer be in a fully polarized state
due to the AF superexchange which favors orbital
alternation. Independent spin, orbital, and coupled spin-orbital
excitations of collective type, as discussed above, therefore can no
longer exist. We will focus, in particular, on the question how spin
excitations are modified by the presence of orbitals in this case.

The paper is organized as follows: In Sec.~\ref{Model} we present a
generalization of the spin-orbital model, Eq.~(\ref{FMSU(4)}), to a model
with anisotropic orbital exchange.
For the extreme quantum limit of the orbital sector interacting via an
XY-type coupling we then derive an effective BF model which resembles
models considered in the context of ultracold BF gases. By using a
density matrix renormalization group algorithm applied to transfer
matrices (TMRG) we exemplarily investigate numerically the crossover
from AF to FM correlations. In Sec.~\ref{MF} we discuss the MF
decoupling approach. We allow for a dimerization in both sectors and
discuss the obtained MF phase diagram. In Sec.~\ref{Sqw_uni} we
summarize the results for the dynamical spin structure factor
$S(q,\omega)$ obtained within the modified spin-wave theory (MSWT) for
the uniform FM spin chain.\cite{MSWT2,MSWT3} In Sec.~\ref{PT} we
formulate an approach where the coupling between spins and orbitals is
treated perturbatively. The approach is based on representing the
spins by bosons using the MSWT\cite{MSWT,MSWTlong} and the orbitals by
Jordan-Wigner fermions.  Finally, in Sec.~\ref{C_PT}, we consider the
effects of coupled spin-orbital degrees of freedom on the
thermodynamics of the system.  We focus, in particular, on the
specific heat as a function of temperature and compare perturbative
results with numerical data obtained by TMRG.  In Sec.~\ref{Conc} we
summarize and discuss our results. The Appendix provides details of
the perturbative approach.

\section{Spin-orbital model and mapping onto a Boson-Fermion model}
\label{Model}

\subsection{One-dimensional spin-orbital model}

We will focus here on the physical situation realized in the vanadium
perovskites, such as YVO$_3$, where the superexchange interactions are
antiferromagnetic. 
In YVO$_3$ the two $d$ electrons occupy the lower lying
$t_{2g}$ orbitals while the $e_g$ orbitals are empty.
From electronic structure
calculations\cite{PhysRevB.54.5368,PhysRevB.58.6831,PhysRevB.60.7309}
it is concluded that the $t_{2g}$ orbitals are split into
a lower-lying $xy$ orbital level and a higher-lying doublet of 
$xz$ and $yz$ orbitals. Therefore the $xy$ orbital will always be
occupied by one electron, controlling the AF correlations
in the $(a,b)$ planes. The large Hund's coupling $J_\textnormal{H}$ (normalized to the interatomic Coloumb interaction $U$) present 
in YVO$_3$ will support parallel alignment of electronic spins at V$^{3+}$ 
ions in $d^2$ configurations,\cite{PhysRevB.61.11506} leading to $S=1$ spins.
Therefore, the remaining electron will be placed in one of the two other 
orbitals $\{xz,yz\}$ which constitutes the $\tau=1/2$ orbital degree of freedom. 
On this basis, an effective spin-orbital superexchange model for YVO$_3$ 
with spins $S=1$ has been derived.\cite{PhysRevLett.86.3879} Here we shall 
study the 1D spin-orbital model extracted from it for the $c$-axis.\cite{SPPRL}
%


The simplest Hamiltonian for the $c$-axis FM chains in YVO$_3$ taking
$J_H$ into account is given by Eq.~(\ref{FMSU(4)}) with $J>0$, $x=1$,
and $y=1/4-\gamma_\text{H}$. Here $\gamma_\textnormal{H}$ is
proportional to Hund's coupling $J_\text{H}$, supporting FM
correlations in the spin sector.
%
For $S=1$ and realistic values of Hund's coupling for vanadates,
$\gamma_\text{H}\sim 0.1$, numerical investigations of this model
showed strong but short-ranged dimer correlations in a certain finite
temperature range caused by the related entropy gain although the
ground state is uniformly FM.\cite{PhysRevB.67.100408} The
same Hamiltonian was also studied using a MF decoupling
scheme.\cite{SPPRL} Within this approach a finite temperature phase
with dimer order in both sectors was found.  However, as discussed in
the introduction, the MF decoupling approach
has severe limitations as it
does not take the coupled spin-orbital dynamics into account. 

We start from a generalization of Eq.~(\ref{FMSU(4)}) which reads
\begin{equation}
\label{HGamma}
\mathcal{H}_{S\tau}(\Gamma)=J\sum_j\left(\vec{S}_j\!\cdot\!\vec{S}_{j+1}+x\right)
\left(\left[\vec{\tau}_j\!\cdot\!\vec{\tau}_{j+1}\right]_\Gamma+y\right)\,,
\end{equation}
with
\begin{equation}
\label{tauGamma}
\left[\vec{\tau}_j\cdot\vec{\tau}_{j+1}\right]_\Gamma\equiv
\vec{\tau}_j\cdot\vec{\tau}_{j+1}-\Gamma\tau_j^z\tau_{j+1}^z\,.
\end{equation}
The calculations presented here are valid for general $S$ and $x$, but
we will, unless stated otherwise, only address the case $S=1$, $x=1$
relevant for YVO$_3$ in the following. For $\Gamma=1$ the pseudospin
sector reduces to an XY model.

\begin{figure}[t!]
\includegraphics*[width=1.0\columnwidth]{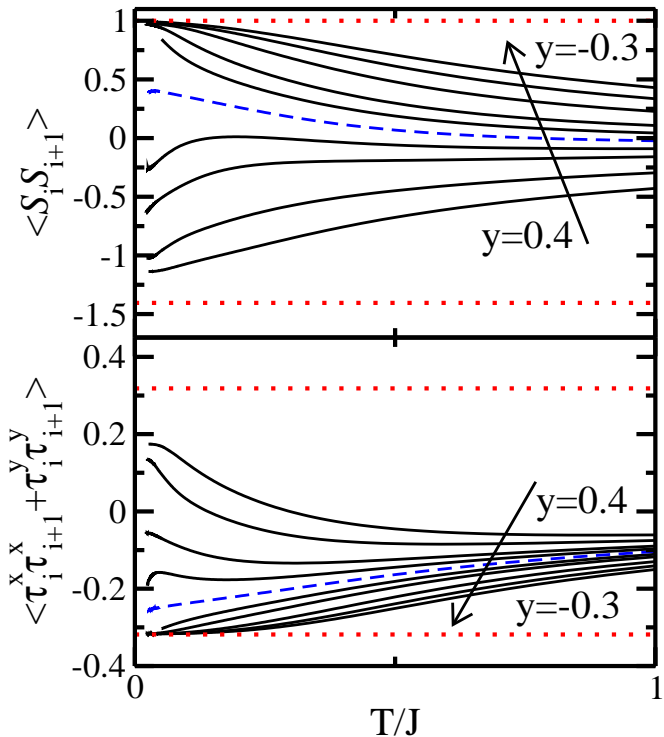}
\caption{(Color online) Nearest-neighbor spin and orbital correlation functions for 
the spin-orbital model (\ref{HGamma}) with $\Gamma=1$ as a
          function of temperature $T$ in units of $J$ (we set $k_\text{B}=1$). In both panels $y=0.4$, 0.3, 0.2,
          0.15, 0.1, 0.05, 0.0, $-0.1$, $-0.2$, $-0.3$ in arrow direction.
          The spin correlations switch from AF to FM at $y=0.1$
          (dashed lines).  The dotted lines in the upper (lower) panel
          correspond to the limiting values $1$ and $-1.4015$ ($-1/\pi$
          and $1/\pi$), respectively.}
\label{Fig_NN_TMRG}
\end{figure}

In Fig.~\ref{Fig_NN_TMRG} the nearest-neigbor spin and orbital
correlation functions for $\Gamma=1$ as a function of temperature 
for various parameters $y$ obtained by TMRG are shown. This method allows 
us to obtain thermodynamic quantities for 1D quantum systems directly
in the thermodynamic
limit.\cite{BursillXiang,WangXiang,SirkerKluemperEPL}
For $y\lesssim 0.1$ the ground state has ferromagnetically aligned spins.
The phase transition between the fully polarized FM state and a state
with AF spin correlations at $y\approx 0.1$ is first order. In the
limit $y\gg 1$ the value $\langle \vec{S}_j\vec{S}_{j+1}\rangle\simeq
-1.4015$ for a Haldane $S=1$ Heisenberg chain is
reached,\cite{WhiteHuse} while the orbital correlations approach $\langle
\tau^x_j\tau^x_{j+1} +\tau^y_j\tau^y_{j+1}\rangle \to 1/\pi$.  We note
that the FM ground state is lost at $y\simeq 0.1$ both in the model
with $\Gamma=1$ investigated here as well as in the model with an
isotropic pseudospin sector
($\Gamma=0$).\cite{Kawakami,PhysRevB.67.100408} For the model with
$\Gamma=1$, however, we have a direct phase transition from the FM to
the Haldane phase while for the isotropic model an orbital valence
bond phase is intervening between these two phases.\cite{Kawakami}

The isotropic model (\ref{HGamma}), $\Gamma=0$, has also been
intensely studied for $S=\tau=1/2$. Here the phase diagram is more
complex than in the $S=1$, $\tau=1/2$ case.\cite{ItoiQin,ChenWang} For
$x=1$ the FM spin state is again found to be stable for
$y\lesssim 0.1$. However, now the transition at $y\sim 0.1$ is to a
gapless ``renormalized $\text{SU}(4)$'' phase followed by a further phase
transition at larger $y$ into a dimer phase. The phase with
ferromagnetically polarized orbitals is absent because
$\langle \vec{S}_j\cdot\vec{S}_{j+1}+x\rangle >0$ for $x=1$ but is
again present for large $y$ if $x\lesssim \ln 2-1/4$.

\subsection{Boson-fermion model}

For the 1D spin-orbital model, Eq.~(\ref{HGamma}), at the point
$\Gamma=1$, we will now derive an effective BF model which will be
used as a starting point for the perturbative approach.  First,
applying the Jordan-Wigner transformation the orbital part is mapped
onto a free fermion model (for $\Gamma\ne1$ the pseudospins map onto
interacting fermions). The spin part of the spin-orbital Hamiltonian
(\ref{HGamma}) will be represented by bosons.
Concentrating on the case 
where the spin part is ferromagnetically polarized in the ground state,
we can treat the spin sector by the MSWT.\cite{MSWT,MSWTlong} To this end,
we introduce bosonic operators by a
Dyson-Maleev transformation.
If we retain bosonic operators only up to quadratic order we end up with
\begin{equation}\label{start}
	H\equiv \mathcal{H}_{S\tau}(1)-JN(S^2+x)y\simeq H_0+H_1\,.
\end{equation}
Here $H_0$ is already diagonal
\begin{equation}
\label{BF0}
	H_0=\sum_k\omega_\text{B}(k)b_k^\dagger b_k^{\ }
+\sum_q\omega_\text{F}(q) f_q^\dagger f_q^{\ }\,,
\end{equation}
with $f_q^ \dagger$ and $f_q^ {\ }$ ($b_k^ \dagger$ and $b_k$) being
the fermionic (bosonic) creation and annihilation operators,
respectively. The magnon dispersion is given by
\begin{equation}\label{Mag}
	\omega_\text{B}(k)=2JS|y|(1-\cos k)\, ,
\end{equation}
and the fermion dispersion reads
\begin{equation}\label{Orb}
	\omega_\text{F}(q)=J(S^2+x)\cos q \,.
\end{equation}
The spinless fermions fill up the Fermi sea between the Fermi points 
at $k_F=\pm\pi/2$. 

For FM spin chains usual spin-wave theory has to be
modified by a Lagrange multiplier $\mu$ acting as a
chemical potential which enforces
the Mermin-Wagner theorem of
vanishing magnetization at finite temperature\cite{MSWT}
\begin{equation}\label{MSWTC1}
	S=\frac{1}{N}\sum_k\left<b_k^\dagger b_k^{\ }\right>.
\end{equation}
Thermodynamic quantities calculated with this method are in excellent
agreement with the exact Bethe ansatz solution for the uniform chain as 
well as with numerical TMRG data for the dimerized FM chain for 
temperatures up to $T\sim |J_\text{eff}|S^2$,
with $J_\text{eff}$ being the effective exchange constant 
of the model under consideration.\cite{SPPRL,MSWT,MSWTlong,ICM} 

The interacting part couples bosons and fermions and reads
\begin{equation}
\label{HC2}
	H_1=\frac{1}{N}\sum_{k_1,k_2,q}
\omega_\text{BF}(k_1,k_2,q)b_{k_1}^\dagger b_{k_2}^{\ }
f_{q}^\dagger f_{k_1-k_2+q}^{\ },
\end{equation}
with the vertex
\begin{equation}
	\begin{aligned}
\omega_\text{BF}(k_1,k_2,q)\equiv &JS\left[\cos(k_2-q)+\cos(k_1+q)\right.\\
&\left.-\cos(k_1-k_2+q)-\cos q\right].
	\end{aligned}
\end{equation}
The Hamiltonian $H=H_0+H_1$ with $H_0$ and $H_1$ given by
Eqs.~(\ref{BF0}) and (\ref{HC2}), supplemented by
the constraint (\ref{MSWTC1}), is an effective BF representation valid
at low temperatures. We will investigate this model in Sec.~\ref{PT}
treating the BF coupling perturbatively.

\section{Mean-field decoupling}
\label{MF}
The spin-orbital model (\ref{HGamma}) contains rich and interesting
physics. A first attempt to understand the properties of the model is
to apply a MF decoupling which neglects the coupled spin-orbital
degrees of freedom and treats the spin-orbital chain as two separate
chains with effective coupling constants which have to be determined
self-consistently. Note, however, that this treatment does not involve
site variables as in the classical Weiss-MF theory but takes the
correlations on a bond as relevant variables. Interestingly, these
expectation values never vanish, which makes them useful particularly
in cases without long-range order.
\subsection{Decoupling into spin and orbital chain}

Applying a MF decoupling and allowing for a dimerization in both
sectors\cite{SPPRL,OlesHorsch} we obtain from Eq.~(\ref{HGamma})
\begin{equation}\label{HSpHtau}
\mathcal{H}_{S\tau}(\Gamma)\simeq H^\text{MF}_S+H^\text{MF}_{\tau}(\Gamma)\,,
\end{equation}
with the spin and orbital Hamiltonians
\begin{equation}\label{MFeq}
	\begin{aligned}
		&H^\text{MF}_S=\mathcal{J}_S\sum_{j=1}^N
\{1+(-1)^j\delta_S\}\vec S_j\cdot\vec S_{j+1}\,,\\
  		&H^\text{MF}_\tau(\Gamma)=\mathcal{J}_\tau\sum_{j=1}^N
\{1+(-1)^j\delta_\tau\}\left[\vec{\tau}_j\cdot\vec{\tau}_{j+1}\right]_{\Gamma}.
	\end{aligned}
\end{equation}
Within this
approximation the effective superexchange constants and dimerization
parameters are given by
\begin{equation}
\label{Selfcons}
	\begin{aligned}
	\mathcal{J}_\tau&=J\,\frac{\Delta^+_{SS}+2x}{2},
&\qquad\delta_\tau=\frac{\Delta_{SS}^-}{\Delta_{SS}^++2x}\,, \\
	\mathcal{J}_S&=J\,\frac{\Delta^+_{\tau\tau}+2y}{2},
&\qquad\delta_S=\frac{\Delta_{\tau\tau}^-}{\Delta_{\tau\tau}^++2y}\,,
	\end{aligned}
\end{equation}
where we have defined
\begin{equation}\label{Deltas}
	\begin{aligned}
		\Delta_{SS}^\pm&=\left<\vec S_{2j}\cdot\vec S_{2j+1}\right>
\pm\left<\vec S_{2j}\cdot\vec S_{2j-1}\right>, \\ 
		\Delta_{\tau\tau}^\pm&=\left<\left[
\vec{\tau}_{2j}\cdot\vec{\tau}_{2j+1}\right]_{\Gamma}\right>\pm\left<
\left[\vec{\tau}_{2j}\cdot\vec{\tau}_{2j-1}\right]_{\Gamma}\right>.\\
	\end{aligned}
\end{equation}
Here $\Delta_{\sigma\sigma}^-$ with $\sigma=S$ ($\sigma=\tau$) is an
order parameter for the spin (orbital) dimerization, respectively.
Thus, the exchange constants and dimerization parameters for each
sector are determined by the nearest-neighbor correlation functions
in the other sector, making a self-consistent calculation
necessary. In the following we want to solve
Eqs.~(\ref{MFeq})-(\ref{Deltas}) for the spin exchange being
effectively FM, i.e. $\mathcal{J}_S<0$.

\subsection{Dimerized orbital correlations}
Numerical investigations of the model with isotropic orbital exchange,
$\mathcal{H}_{S\tau}(0)$, have shown orbital-singlet formation in the
ground state\cite{Kawakami} for $y\gtrsim0.1$. Moreover, although the
ground state consists of a fully spin polarized FM state for
$y\lesssim0.1$ with AF orbital correlations, it has been shown that a
tendency towards orbital singlet formation is still present but has to
be activated by thermal fluctuations.\cite{PhysRevB.67.100408} In
Ref.~\onlinecite{SPPRL} the model (\ref{HSpHtau}) was studied in the
FM regime with $x=1$, $y=\frac{1}{4}-\gamma_\textnormal{H}$,
$\Gamma=0$ and $\gamma_\textnormal{H}=0.1$ in order to address the
question whether this orbital-Peierls effect can be captured within a
MF decoupling approach. A dimerized phase for $0.10\lesssim
T/J\lesssim0.49$ (we set $k_\text{B}=\hbar=1$) was found with the dimerization amplitude in the spin
sector being much larger than in the orbital sector.

We now want to compare this result with the case where we set
$\Gamma=1$ in Eq.~(\ref{HSpHtau}) so that the self-consistent
Eqs.~(\ref{Selfcons}) can be solved analytically by applying a Jordan-Wigner transformation and
MSWT. Introducing fermionic operators $f^{(\dagger)}_{j,e}$ if the
index $j$ is even and $f^{(\dagger)}_{j,o}$ if $j$ is odd for the
pseudospins, we rewrite $H^\text{MF}_\tau\equiv H^\text{MF}_\tau(\Gamma=1)$
in Fourier representation. Finally introducing new fermionic operators
$\phi^{(\dagger)}_q$ and $\varphi^{(\dagger)}_q$ which
diagonalize the Hamiltonian $H^\text{MF}_\tau$, we find
\begin{equation}
H^\text{MF}_\tau=\sum_q\omega^{\text{MF}}_\text{F}(q,\delta_\tau)
(\phi_q^\dagger\phi_q^{\
}+\varphi_q^\dagger\varphi_q^{\ })\,,
\end{equation}
with the fermionic dispersion\cite{Pincus19711971} 
\begin{equation}\label{disfermMF}
\omega^{\text{MF}}_\text{F}(q,\delta_\tau)\equiv
\mathcal{J}_\tau\sqrt{\cos^2q+\delta_\tau^2\sin^2q}\,.
\end{equation}

We can now calculate $\Delta_{\tau\tau}^\pm$, as given in Eq.~(\ref{Deltas}) straightforwardly
and obtain
\begin{equation}
\begin{aligned}\label{Deltataum}		\Delta_{\tau\tau}^-&=&\!\!\!\frac{2\delta_\tau}{N}\sum_q
\frac{\left\{2n_\textnormal{F}[\omega^\text{MF}_\text{F}(q,\delta_\tau)]-1\right\}\sin^2q}{\sqrt{\cos^2q+\delta_\tau^2\sin^2q}}
, \\
\Delta_{\tau\tau}^+&=&\!\!\!\frac{2}{N}\sum_q
\frac{\left\{2n_\textnormal{F}[\omega^\text{MF}_\text{F}(q,\delta_\tau)]-1\right\}\cos^2q}{\sqrt{\cos^2q+\delta_\tau^2\sin^2q}}
,
\end{aligned}
\end{equation}
where $n_\textnormal{F}(x)=\left\{\exp(\beta x)+1\right\}^{-1}$ is the
Fermi function and $\beta=1/T$.

\subsection{Dimerized spin correlations}

Next we turn to the spin part of Eq.~(\ref{HSpHtau}) to which we apply the
MSWT.\cite{MSWT,MSWTlong} We introduce two bosonic operators
$b^{(\dagger)}_{j,e}$ [$b^{(\dagger)}_{j,o}$] for $j$ even [odd] by
means of a Dyson-Maleev transformation. Retaining only terms bilinear
in the bosonic operators we can diagonalize the resulting Hamiltonian
by a Bogoliubov
transformation 
leading to
\begin{equation}
	\begin{aligned}
H^\text{MF}_S=&\sum_k\left\{\omega^{\text{MF}}_{\text{B},-}(k,\delta_S)
\alpha_k^\dagger\alpha_k^{\
}+\omega^{\text{MF}}_{\text{B},+}(k,\delta_S)\beta_k^\dagger\beta_k^{\ }\right\}\\
&+\mathcal{J}_SNS^2\,,
	\end{aligned}
\end{equation}
with
the two magnon branches
\begin{equation}
	\omega_{\text{B},\pm}^{\text{MF}}(k,\delta_S)=2|\mathcal{J}_S|S
\left(1\pm\sqrt{\cos^2k+\delta_S^2\sin^2k}\right)\,.
\end{equation}
The constraint of vanishing magnetization at finite temperature
(\ref{MSWTC1}) now reads
\begin{equation}
\label{MSWTconstraint}
	S=\frac{1}{N}\sum_k\left\{n_\textnormal{B}[\zeta^-(k,\delta_S)]
+n_\textnormal{B}[\zeta^+(k,\delta_S)]\right\},
\end{equation}
where $n_\textnormal{B}(x)=\{\exp\{\beta x\}-1\}^{-1}$ is the Bose
function and
$\zeta^\pm(k,\delta_S)=\omega^\text{MF}_{\textnormal{B},\pm}(k,\delta_S)-\mu(\delta_S)$.

\begin{figure}[t!]
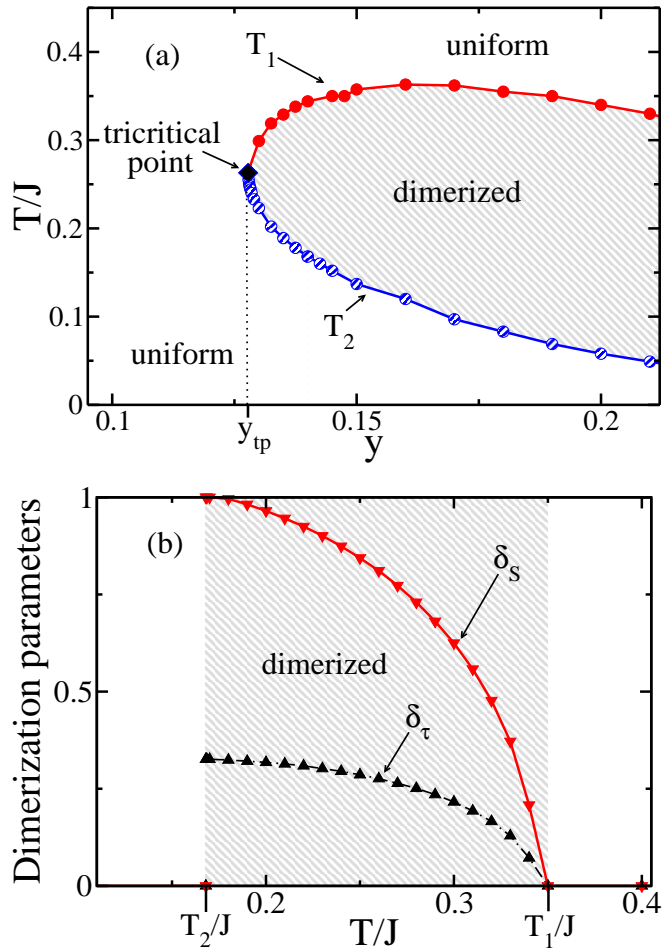

	\includegraphics[width=.99\linewidth,angle=-0]{DimPD.eps}\\\vspace*{.25cm}
	\includegraphics[width=.99\linewidth]{DimSing.eps}
	\caption{(a) Phase diagram of the Hamiltonian (\ref{HSpHtau})
          with $\Gamma=1$ and $x=1$ in mean-field decoupling. The
          shaded area represents the dimerized phase. The phase
          transition at $T_2$ is first order whereas the transition at
          $T_1$ is of second order. The two transition lines merge at
          the tricritical point $y_\text{tp}$. (b) Dimerization
          parameters $\delta_S$ and $\delta_\tau$ for $x=1$ and
          $y=0.14$.  The lines are guides to the eye. The shaded area
          marks the temperature range where the dimerization is
          nonzero.}\label{SimpleMF}
\end{figure}

To calculate the nearest-neighbor correlation functions $B_\pm\equiv\bigl<\vec S_j\cdot\vec
S_{j\pm1}\bigr>$ it is necessary to go beyond linear spin-wave theory.
Taking terms of quartic order into account and using
Eq.~(\ref{MSWTconstraint}) we obtain\cite{SPPRL}
\begin{equation}\label{MSWTCFe}
B_\pm=\left(\frac{1}{N}\sum_kf_\pm(k,\delta_S)\sum_{\sigma\in\{\pm\}}\sigma
 n_\textnormal{B}[\zeta^\sigma_B(k,\delta_S)]\right)^2.
\end{equation}
Here we have defined
\begin{equation}
\label{fpm}
f_\pm(k,\delta_S)\equiv\frac{\cos^2 k\pm \delta_S\sin^2 k}
{\sqrt{\cos^2k+\delta_S^2\sin^2k}} \, .
\end{equation}
From these expressions we can obtain $\Delta_{SS}^\pm$ which, combined
with Eq.~(\ref{Deltataum}), allows us to solve
Eqs.~(\ref{HSpHtau})-(\ref{Deltas}) self-consistently. 

\subsection{Mean-field phase diagram}
We first discuss the ground state phase diagram of the Hamiltonian
(\ref{HSpHtau}) for $\Gamma=1$. Depending on the sign of the
effective coupling constant $\mathcal{J}_S$ we find
$\langle\vec{S}_j\vec{S}_{j+1}\rangle =1,\, -1.4015$ with the latter
value being the approximate result for the $S=1$ AF Haldane chain. In
the following, we restrict our discussion to $-1<x<1.4015$ so that
$\langle\vec{S}_j\vec{S}_{j+1}\rangle$ and $\mathcal{J}_\tau
=J(\langle\vec{S}_j\vec{S}_{j+1}\rangle + x)$ always have the same
sign. For the orbital sector we obtain, on the other hand, $\langle
\tau^x_j\tau^x_{j+1}+\tau^y_j\tau^y_{j+1}\rangle =\pm 1/\pi$.
$y>1/\pi$ implies $\mathcal{J}_S>0$ and the ground state is therefore
certainly AF (Haldane phase) whereas $\mathcal{J}_S<0$ for $y<-1/\pi$
leading to a FM state. In the regime $-1/\pi<y<1/\pi$ the
self-consistent equations have two solutions with energies
$E_0^\text{AF}\approx (1/\pi+y)(-1.4015+x)$ and
$E_0^\text{FM}=(-1/\pi+y)(1+x)$ and a first order phase transition
between the FM and AF states occurs where the energies cross. For the
case $x=1$ we are focussing on here, this happens at $y_c\approx
0.212$ and the FM state is stable for $y<y_c$.  Compared to the
numerical solution where $y_c\approx 0.1$ (see Fig.~\ref{Fig_NN_TMRG})
the range of stability of the FM state is therefore increased in the
MF solution.

Next, we investigate the possibility of a finite temperature
dimerization for $x=1$ in that part of the phase diagram where the
ground state is FM.
As shown in Fig.~\ref{SimpleMF}(a) we find that a dimerized phase at
finite temperatures does indeed exist in MF decoupling for
$y_\text{tp}\approx0.128\lesssim y\lesssim y_c\approx0.212$ where $y_\text{tp}$ denotes the tricrictal point. As in the model with an isotropic
pseudospin sector,\cite{SPPRL} the temperature range where the
dimerized phase is stable depends on $y$.  At the onset temperature $T_1$ the phase transition is of second order whereas at the reentrance temperature $T_2$ it is of first order, see Fig.~\ref{SimpleMF}(b). As in the case
$\Gamma=0$, the dimerization in the spin sector is always much larger
than in the orbital sector.


As pointed out before, the MF decoupling suffers from severe limitations and it is expected to be an even worse
approximation in the extreme quantum case $\Gamma = 1$ than in the
case $\Gamma=0$ studied previously.\cite{SPPRL} In particular the coupling between spin and orbital degrees of freedom is completely lost within this approach. In the following
sections we will therefore develop an alternative
perturbative treatment of the spin-orbital coupling.

\section{Dynamical spin structure factor 
for the uniform ferromagnetic chain}
\label{Sqw_uni}

\begin{figure}[t!]
\includegraphics[width=.95\linewidth]{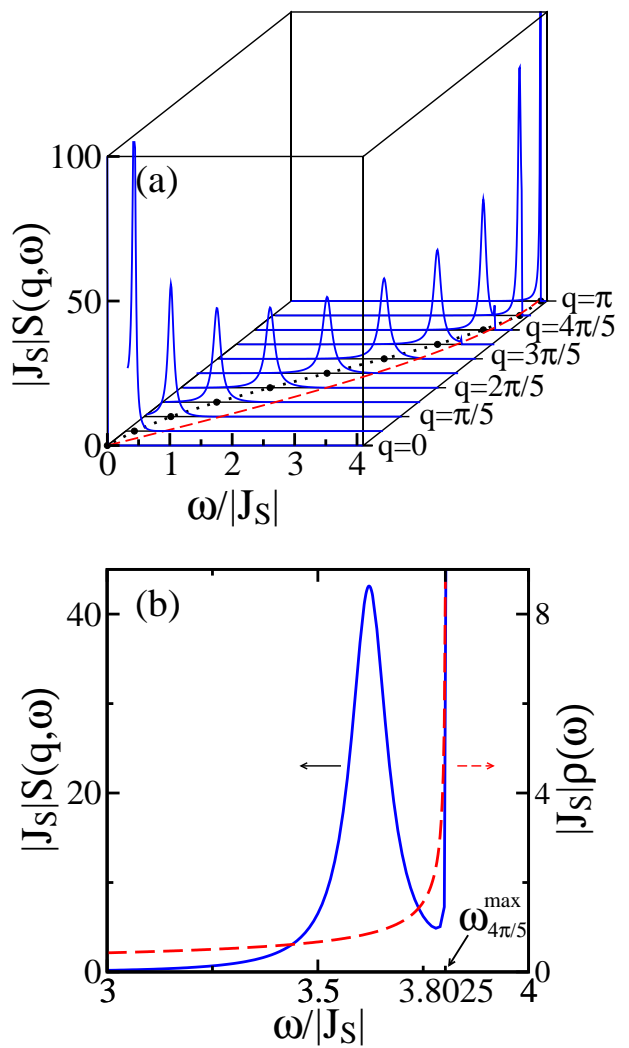}
\caption{(Color online)
(a) Dynamical spin structure factor $S(q,\omega)$ as obtained for 
$T/|J_S|=0.1$ and $0\le q\le\pi$. The dashed line indicates the
upper boundary of the two magnon continuum. The dots are 
projections of the peak positions onto the $(q,\omega)$ plane. 
They are connected by the dotted line which is a guide to the eye. 
(b) Dynamical spin structure factor $S(q,\omega)$ for the same
          parameters at $q=4\pi/5$ (solid line) and the
          corresponding density of states (dashed line). 
}
\label{-2ImGo}
\end{figure}
In order to investigate coupled spin-orbital degrees of freedom and,
in particular, their implications on the spin dynamics of the
spin-orbital chain, a detailed understanding of the spin dynamics of a
FM chain is useful. We shall avoid the complications of the dimerized
chain and focus our study on the uniform 1D ferromagnet.\cite{remarkdim} In doing so
we neglect the coupling between spin and pseudospin operators for a
moment and consider
\begin{equation}\label{HS}
H_S=J_S\sum_j\vec S_j\cdot\vec S_{j+1}\,,
\end{equation}
with $J_S<0$. It is
well-known that MSWT does not respect the SU(2) symmetry of the FM
Heisenberg chain Eq. (\ref{HS}). We therefore directly calculate the 
full spin correlation function\cite{MSWT2,MSWT3}
\begin{equation}\label{G0rt}
G(r,\tau)\equiv-\left<\mathcal{T}[
\textbf{S}_j(0)\cdot\textbf{S}_{j+r}(\tau)]\right> \; .
\end{equation}
In Fourier space we obtain
\begin{equation}\label{G0qomega}
 G(q,\omega_{\nu,\textnormal{B}})\!=\frac{1}{N}\sum_k
(1+n_\textnormal{B}[\zeta(k)])n_\textnormal{B}[\zeta(q-k)]
\frac{1-\textnormal{e}^{-\beta\epsilon_q(k)}}
{i\omega_{\nu,\textnormal{B}}-\epsilon_q(k)},
\end{equation}
where we have used the bosonic Matsubara frequencies
$\omega_{\nu,\textnormal{B}}$ and
$\epsilon_q(k)\equiv\zeta(k)-\zeta(q-k)$ with $k\in[-\pi,\pi]$. 
The reduced magnon dispersion reads
\begin{equation}\label{Magnons}
	\zeta(k)=2J_SS(1-\cos k)-\mu.
\end{equation}

In Fig.~\ref{-2ImGo}(a) the dynamical spin structure factor,
\begin{equation}\label{DSSFFM}
S(q,\omega)=2n_\text{B}(-\omega)\,\textnormal{Im}G^\textnormal{ret}(q,\omega),
\end{equation}
is shown for the uniform FM chain at $\omega>0$, where
\begin{equation}
\begin{aligned}
\textnormal{Im}G^\textnormal{ret}(q,\omega)=
&\frac{\pi}{N}\sum_k(1+n_\textnormal{B}[\zeta(k)])n_\textnormal{B}[\zeta(q-k)]\\
&\times\left(\textnormal{e}^{-\beta\epsilon_q(k)}-1\right)
\delta(\omega-\epsilon_q(k))
\end{aligned}
\end{equation}
is the imaginary part of the retarded Green's function obtained from
Eq.~(\ref{G0qomega}) by analytical continuation.  Up to a factor of
$2\pi$, as a matter of definition, we obtain the result previously
given by Takahashi.\cite{MSWT2} The structure factor fulfills detailed
balance, $S(q,\omega)=\textnormal{e}^{\beta\omega}S(q,-\omega)$.  The
symbols in Fig.~\ref{-2ImGo} show the peak positions projected onto
the $(q,\omega)$ plane. They follow the reduced dispersion
Eq.~(\ref{Magnons}). Also shown in Fig.~\ref{-2ImGo}(a) as a dashed
curve is $\omega_q^\textnormal{max}=4|J_S|S\sin\frac{q}{2}$
corresponding to the upper boundary of the two magnon continuum
$\epsilon_q(k)$ above which $S(q,\omega)$ is zero in this
approximation.

At the edge of the two magnon continuum $S(q,\omega)$ has a
singularity.  In Fig.~\ref{-2ImGo}(b) the dynamical spin structure
factor for the same parameters as used in Fig.~\ref{-2ImGo}(a) is
shown at $q=4\pi/5$ together with the density of states which is given
by $\rho_q(\omega)=1/\sqrt{(\omega_q^\textnormal{max})^2-\omega^2}$.
Right below the singularity at $\omega_q^\textnormal{max}$ the density
of states to lowest order reads
$\rho_q(\omega_q^\textnormal{max}-\delta\omega)\sim
1/\sqrt{\delta\omega}$, i.e., $S(q,\omega)$ shows a square root
divergence at the upper threshold. If the edge singularity and the
central peak are well separated then the spectral weight of the edge
singularity is much smaller than the spectral weight of the central
peak. If, on the other hand, the edge singularity is close to the
central peak then the shape of the latter is strongly affected by the
occurence of the edge singularity. In this case the edge singularity
gives a significant contribution.

It is instructive to analyze $S(q,\omega)$ in the limit of small $q$. If the edge singularity and the peak of the structure factor are well
separated, the lineshape of the peak can be obtained approximately. To
this end, for small $q$ but $|J_S|S^2q/T\gg1$ we only retain the leading terms of
Eq.~(\ref{G0qomega}). Performing a saddle point approximation
to lowest order we find $S(q,\omega)\sim n_\text{B}(-\omega)(a(q,\omega)-a(q,-\omega))$, with
\begin{equation}
\label{aqomega}
a(q,\omega)\approx2S\frac{\frac{|J_S|Sq}{\xi}}
{(\omega-J_SSq^2)^2+\left(\frac{J_SSq}{\xi}\right)^2}\,.
\end{equation}
This Lorentzian lineshape is only valid for low temperatures. Here $\xi\approx |J_S|S^2/T$ is the correlation
length in the low-temperature limit.\cite{MSWT,MSWTlong,MSWT2,MSWT3} Finally, we want to stress that
for $T\to 0$ the peaks will reduce to $\delta$-functions, i.e., only
thermal broadening is included in this approximation.

\section{Perturbation theory}
\label{PT}
\begin{figure}[t!]
         \vspace*{-.8cm}
\includegraphics*[angle=-0,width=.9\linewidth]{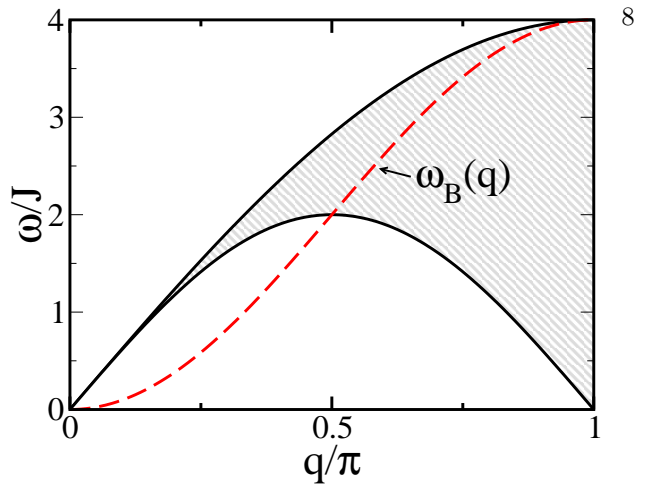}
\caption{(Color online) Magnon dispersion $\omega_\text{B}(q)$ (dashed line) for $y=1$ and
            fermionic two-particle continuum (shaded area).}
\label{FermCont}
\end{figure}

In this section we intent to go beyond the MF decoupling approach treating the influence of the BF interaction, Eq~(\ref{HC2}), on the spin-wave dispersion perturbatively. Naively one would expect that the magnon should be able to couple to the fermionic degrees of freedom if it lies inside the fermionic two-particle continuum. The upper and lower boundary of the latter are given by
\begin{eqnarray}
\label{Continuum}
	\epsilon_\text{F}^\text{max}(q)&= & 2J(S^2+x)\sin(q/2),\nonumber \\
	\epsilon_\text{F}^\text{min}(q)&=& J(S^2+x)\sin q ,
\end{eqnarray}
respectively. The continuum and the magnon dispersion
$\omega_\text{B}(q)$ for $y=-1$ are shown in Fig.~\ref{FermCont}. One
would therefore expect that in this case $\omega_\text{B}(q)$ is
unaffected by the presence of the fermions for $q<\pi/2$, since it can
not couple to these degrees of freedom. However for higher momenta the
spin wave may couple to the fermionic degrees of freedom and thus a
broadening of $S(q,\omega)$ should occur. Moreover by choosing
different values for $y$ the point at which the magnon enters the
fermionic two-particle continuum is changed. Thus the momentum at
which the spin wave is affected by the coupling to the fermionic
degrees of freedom depends directly on the parameter $y$.
\begin{figure*}[t!]
\includegraphics[width=.88\linewidth]{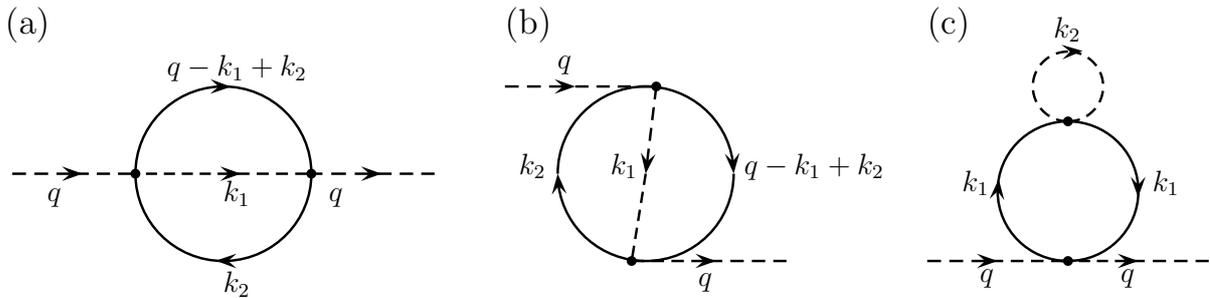}
\caption{Diagrams which contribute to a renormalization of the magnon
  in a perturbation theory: (a,b) Diagrams with momentum exchange
  between the magnon and the fermions, and (c) diagram without
  momentum exchange. All diagrams are second order.  Fermionic
  propagators are shown by solid lines, whereas bosonic propagators
  are shown as dashed lines.  }
\label{Feyn}
\end{figure*}

These arguments give the qualitatively correct picture, i.e., we find
indeed that the coupling of the magnon to the fermionic degrees of
freedom has strong effects on the dynamical spin structure factor at
intermediate and high momenta and that the onset of these effects can
be well estimated by our simple argument. However, there are also
certain aspects which can not be captured within this picture.
For instance, for $2S|y|>(S^2+x)$ it suggests that the spin wave may
leave the fermionic two-particle continuum at a certain momentum $q_l$
and thus should be unaffected by the BF coupling for
$q>q_l$. The detailed calculation, however, reveals that this is not
true because the spin wave decays into a fermionic particle-hole {\it
  and} a remaining spin wave as will become clear in the following.


\subsection{General formulation}\label{GenFo}

Here we want to study the Hamiltonian 
$H$ given by Eq. (\ref{start}), with its noninteracting part $H_0$ and
interacting part $H_1$ defined by Eqs. (\ref{BF0}) and (\ref{HC2}),
by treating the BF interaction perturbatively, i.e., in the limit
$|x|,|y|\gg 1$. As explained in the appendix, we start by performing a
MF decoupling for the interaction $H_1$. Corrections to this solution
are then taken into account perturbatively. Here we adress the bosonic
Green's function at zero temperature
\begin{equation}\label{gb0}
\mathcal{G}_\textnormal{B}^{\ }(q,t)=-i\left<\mathcal{T}_t
\left[b_q^{\ }(t)b_q^\dagger(0)\right]\right>.
\end{equation}
In Fig.~\ref{Feyn} all distinct, connected diagrams beyond the MF decoupling up to second order
are shown. 

We calculate the Green's function from the Dyson equation
\begin{equation}\label{Dy}
\mathcal{G}_\text{B}(q,\omega)=
\frac{1}{\left\{\mathcal{G}_\text{B}^{(0)}(q,\omega)\right\}^{-1}-\Sigma(q,\omega)},
\end{equation}
with
\begin{equation}
\left\{\mathcal{G}^{(0)}_\text{B}(q,\omega)\right\}^{-1}=\omega-\zeta(q),
\end{equation}
where $\zeta(q)$ is the reduced magnon dispersion defined in
Eq.~(\ref{Magnons}) with $J_S=J(y-1/\pi)$, and the self-energy
$\Sigma(q,\omega)$ is approximated by the proper self-energy
$\Sigma_2(q,\omega)$ obtained by summing up the diagrams which can be
composed of the diagrams shown in Fig.~\ref{Feyn}.

\begin{figure}[t!]
\includegraphics[width=.88\linewidth]{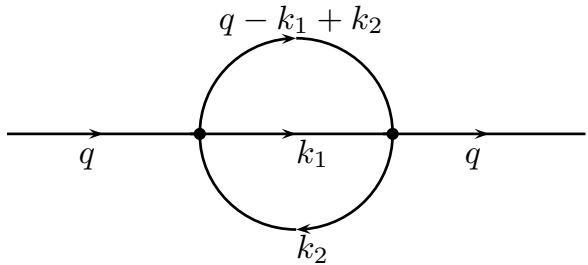}
\caption{Second order diagram for a system of interacting fermions with momentum exchange.}
\label{Fermion}
\end{figure}

\begin{figure*}[t]
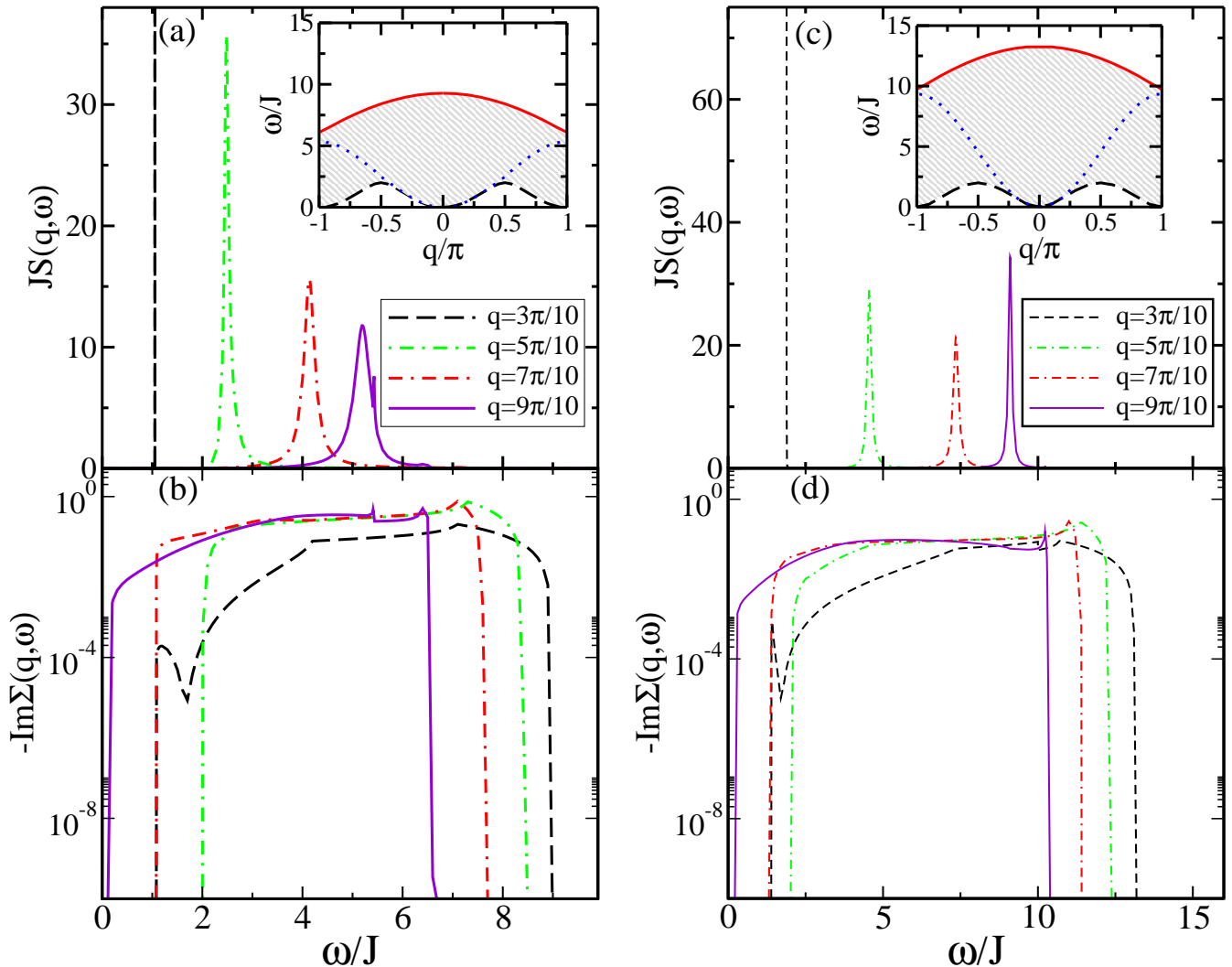

\begin{minipage}{.48\linewidth}
  \includegraphics[width=.99\linewidth]{HFSODSSFym1.eps}
\end{minipage}
\hspace*{.01\linewidth}
\begin{minipage}{.48\linewidth}
  \includegraphics[width=.99\linewidth]{HFSODSSFym2.eps}
\end{minipage}
\caption{(Color online) Perturbative results for the BF model at zero
  temperature with $x=1$.  In the left (right) panel $y=-1$ ($y=-2$),
  respectively. (a),(c) $S(q,\omega)$ with the inset showing the
  region for which $\text{Im}\,\Sigma_\text{BF}^+(q,\omega)$ is
  nonzero as a shaded area and the renormalized spin-wave dispersion
  as a dotted line.  While $S(q,\omega)$ is sharply peaked at low
  momenta, a significant broadening occurs at higher $q$. Moreover we
  find additional structures which, as explained in the text, are due
  to coupled spin-orbital excitations. (b),(d)
  $-\text{Im}\,\Sigma_\text{BF}^+(q,\omega)$ as given in
  Eq.~(\ref{sigma22}) for the corresponding values of $q$ shown in (a)
  and (c), respectively. The coupled spin-orbital excitations show up
  as peaks and edges in $\text{Im}\,\Sigma_\text{BF}^+(q,\omega)$
  (notice the logarithmic scale).}
\label{BFCymp75}
\end{figure*}

The diagrams shown in Figs.~\ref{Feyn}(a) and \ref{Feyn}(b) are of
particular interest because they are the lowest order diagrams where
bosons and fermions exchange momentum.
They describe the part of spin-orbital dynamics which cannot be
captured within the MF decoupling approach discussed in section
\ref{MF}. The diagram shown in Fig.~\ref{Feyn}(b) has to be
thermally activated, i.e., it does not give any contribution at zero
temperature. The same is true for the diagram shown in
Fig.~\ref{Feyn}(c).  Thus at $T=0$ the only second order diagram which
contributes to the self energy is the one shown in
Fig.~\ref{Feyn}(a) leading to 
\begin{equation}\label{sigma22}
\begin{aligned}
\Sigma_\text{BF}^+(q,\omega)=&-\frac{1}{N^2}\sum_{k_1,k_2}\omega^2_\text{BF}(q,k_1,k_2)\\
&\times\frac{\Theta[\omega_\text{F}(q-k_1+k_2)]
\Theta[-\omega_\text{F}(k_2)]}{\omega-\Omega^+_q(k_1,k_2)+i0^+}\,,
\end{aligned}
\end{equation}
where we have abbreviated\cite{FT0}
\begin{equation}
\label{Omega_q}
\Omega^\pm_q(k_1,k_2)\equiv
\pm\zeta(k_1)+\omega_\text{F}(q-k_1+k_2)-\omega_\text{F}(k_2)\,,
\end{equation}
with $k_1,k_2\in[-\pi,\pi]$. 

For systems of interacting fermions we know that perturbation theory
in one dimension often leads to infrared
divergencies.\cite{Giamarchi,Metzner} Such divergencies occur, for
example, for the fermionic analogon of the diagram with momentum
exchange, see Fig.~\ref{Fermion}. These problems can be overcome by
the Dzyaloshinski-Larkin solution or bosonization techniques. For the
model considered here, however, we find no divergencies within the
considered diagrams. One reason for this behavior is a lack of
nesting. While for a fermionic interaction as shown in
Fig.~\ref{Fermion} all the dispersions in the denominator of Eq.~(\ref{sigma22}) are approximately linear at low energies here one of
the dispersions is approximately quadratic so that nesting only occurs
for singular points.  As a further check, we have evaluated the
integrals in Eq.~(\ref{sigma22}) for a constant vertex at small $q$
and $\omega$ and did not find any infrared divergencies.

For finite temperatures the Matsubara formalism can be applied
straightforwardly. The self-energy Eq.~(\ref{sigma22}) now reads 
\begin{eqnarray}\label{SigmaBF}
\Sigma_\text{BF}^+(q,\omega_{\nu,\textnormal{B}})&=& -\frac{1}{N^2}\sum_{k_1,k_2}\frac{\omega^2_\text{BF}(q,k_1,k_2)}
{i\omega_{\nu,\textnormal{B}}-\Omega^+_q(k_1,k_2)}\nonumber \\
&\!\!\!\!\!\!\!\!\!\!\!\!\!\!\!\!\times& \!\!\!\!\!\!\!\!\!\!\!\!
\mathcal{N}^+_\textnormal{F,B}(k_1,k_2,\omega_{\nu,\textnormal{B}},T)
\mathcal{N}_\textnormal{F,F}(q,k_1,k_2,\omega_{\nu,\textnormal{B}},T), \nonumber \\
\end{eqnarray}
where we have abbreviated
\begin{eqnarray}\label{NFNB}
\mathcal{N}^\pm_\textnormal{F,B}(k_1,k_2,\omega_{\nu,\textnormal{B}},T)\!\!&\equiv&\!\!
n_\textnormal{B}[\zeta(k_1)]\pm n_\textnormal{F}[\omega_\text{F}(k_2)],  \nonumber \\
\!\!\!\mathcal{N}_\textnormal{F,F}(q,k_1,k_2,\omega_{\nu,\textnormal{B}},T)\!\!&\equiv&\!\! n_\textnormal{F}[\omega_\text{F}(q-k_1+k_2)]  \nonumber \\
&& -n_\textnormal{F}[\omega_\text{F}(k_2)-\zeta(k_1)]. 
\end{eqnarray}
Note that at finite temperatures both, the reduced spin-wave dispersion $\zeta(q)$ as well as the fermionic dispersion is renormalized due to the MF decoupling applied to Eqs.~(\ref{HC2}). The respective expressions are given in the appendix, see Eqs.~(\ref{remag}) and (\ref{refer}).

At finite temperatures also the diagram shown in Fig.~\ref{Feyn}(b)
contributes and is given by
\begin{equation}\label{SigmaBFm}
	\begin{aligned}
		&\Sigma_\text{BF}^-(q,\omega_{\nu,\text{B}})=-\frac{1}{N^2}\sum_{k_1,k_2}\frac{\omega_\text{BF}^2(q,k_1,k_2)}{i\omega_{\nu,\text{B}}-\Omega^{-}_q(k_1,k_2)}\\
		&\times(1+\mathcal{N}^-_\textnormal{F,B}(k_1,k_2,\omega_{\nu,\textnormal{B}},T))\mathcal{N}_\textnormal{F,F}(q,k_1,k_2,\omega_{\nu,\textnormal{B}},T).
	\end{aligned}
\end{equation}

\subsection{Dynamical spin structure factor}
\label{DSSF} 

Below we present the results obtained by
summing up the diagrams shown in Fig.~\ref{Feyn} in a Dyson series, but
replacing the external legs by the SU(2) symmetric function given in Eq. 
(\ref{G0qomega}). While the perturbative results can, strictly speaking, 
only be valid for $|x|,|y|\gg 1$ we extend the results here to more 
physical values $|x|,|y|\sim \mathcal{O}(1)$ where we still expect 
perturbation theory to give at least a qualitatively correct picture.
Numerical results obtained for the dynamical spin structure factor 
within this perturbative approach are shown in Fig.~\ref{BFCymp75} for 
$T=0$ and $y=-1$ and $y=-2$.
In both cases $S(q,\omega)$ is sharply peaked at small momenta whereas
a significant broadening occurs at higher momenta. Note, that within the
MSWT $S(q,\omega)$ is always a $\delta$-function for the pure spin model 
at $T=0$, i.e., the broadening here is solely due to the coupling to
orbital excitations. 

\begin{figure}[t!]
\includegraphics*[angle=0,scale=.325]{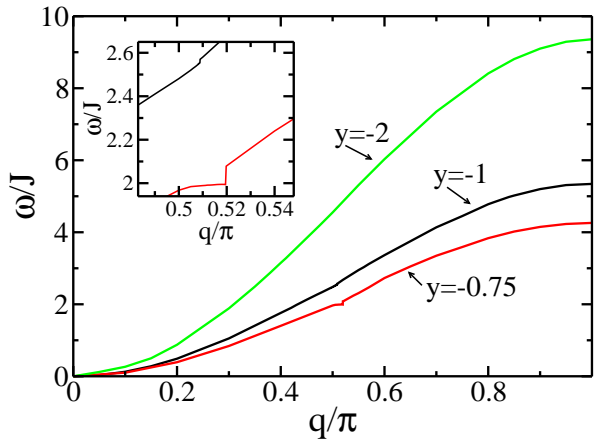}
\caption{(Color online) Renormalized magnon dispersions $\omega_q$ in
  the FM chain for $x=1$ and selected values of $y$. Inset: The most
  pronounced Kohn anomalies occur at $q$ near $\pi/2$.}
\label{RSW}
\end{figure}

By extracting the central peaks of the dynamical spin structure factor
at various momenta, we obtain the renormalized spin-wave dispersion
$\omega_q$ within the perturbative approach. The result of this is
shown in the insets of Fig.~\ref{BFCymp75}(a,c) and in more detail in
Fig.~\ref{RSW}. The magnon dispersion is renormalized and small kinks
are visible close to $q=\pi/2$, which may be interpreted as Kohn
anomalies (see below). The inset of Fig.~\ref{RSW} shows the Kohn
anomalies with a higher resolution. For itinerant ferromagnets Kohn
anomalies are well-known. Here the interaction between the spins of
localized ions is mediated by an exchange with the conduction
electrons.\cite{Kohn,Woll,Barnea,PhysRevLett.16.737,Halilov,Pajda,Moran}
These Kohn anomalies can thus be used to gain information about the
Fermi surface of the conduction electrons.\cite{Kohn,Woll,Barnea}
However to the best of our knowledge Kohn anomalies in the spin-wave
dispersion for insulating materials have not been adressed so far.  As
we will show below, the Kohn anomalies in our case are caused by
coupled spin-orbital degrees of freedom.

Apart from extracting the effective spin-wave dispersion from $S(q,\omega)$, we 
also want to discuss the magnon bandwidth (full width at half maximum (FWHM)) $\Gamma_q$ of the 
central peaks. A broadening of the zero temperature peaks occurs
whenever the imaginary part of the self-energy,
\begin{eqnarray}\label{Imsig}
\text{Im}\ \Sigma_\text{BF}^+(q,\omega)&=&-\frac{\pi}{N^2}\sum_{k_1,k_2}
\omega^2_\text{BF}(q,k_1,k_2)\Theta[-\omega_\text{F}(k_2)]\nonumber \\
&\times&\Theta[\omega_\text{F}(q-k_1+k_2)]\delta(\omega-\Omega^+_q(k_1,k_2)),\nonumber\\
\end{eqnarray}
is non-zero at $\omega=\Omega^+_q(k_1,k_2)$. The contributions within 
the sums are now determined by
the argument of the $\delta$-function as well as by the constraints
given by the Heaviside functions. This procedure, for a given set of 
parameters $x,y,$ and $S$, effectively yields a region
within which the spin wave may scatter on fermion pairs.
Henceforth we call this region the BF continuum. 
The BF continuum is shown in the insets of Figs.~\ref{BFCymp75}(a) and 
\ref{BFCymp75}(c) as shaded areas. The upper
boundary of the BF continuum (solid lines), which is periodic with a period of
$2\pi$, is given by $-4JS(y-1/\pi)+2J(S^2+x)$ for $q=0$, 
and decreases monotonously from this value with increasing $|q|$. The 
lower boundary (dashed lines) is periodic with a period of $\pi$.
\begin{figure}[t!]
	  \includegraphics*[scale=.325]{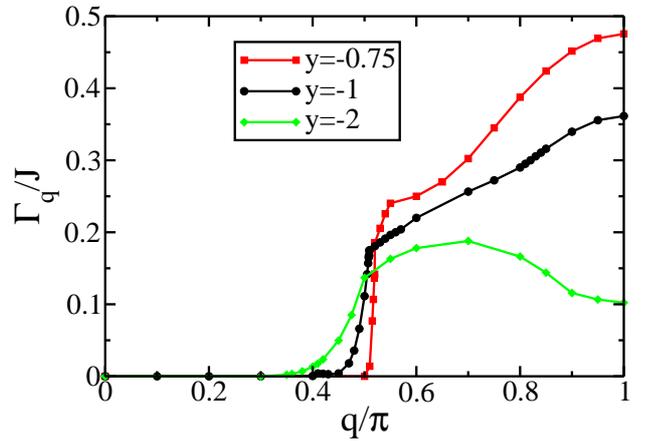}
\caption{(Color online) Magnon linewidth $\Gamma_q$ (FWHM of $S(q,\omega)$) at zero 
temperature (data points), as obtained for $x=1$ and the representative 
values of $y$ indicated in the plot. The lines are guides to the eye.}
\label{BandW}
\end{figure}

To obtain the FWHM of the structure factor, the magnitude of the
contributions to the sums given in Eq.  (\ref{Imsig}) are essential.
Here not only the $(k_1,k_2)$-region which contributes to the
summation but also the magnitude of the vertex $\omega_\text{BF}$ is
of importance. We observe that the vertex is small at small momenta
but increases at intermediate and high momenta. This leads to a strong
increase of the magnitude of the imaginary part of the self-energy as
shown in panels (b) and (d) of Fig.~\ref{BFCymp75}. From the insets of
Fig.~\ref{BFCymp75} it becomes clear that the spin-wave dispersion enters the BF continuum
depending on $y$. For higher values of $|y|$ the spin wave enters at
lower momenta. However, since the vertex gives smaller contributions at
smaller momenta, the broadening of the central peaks of the dynamical
spin structure factor turns out to be smaller the smaller the momenta
are at which the spin wave enters the BF continuum. 
This can be seen in Fig.~\ref{BandW} where the magnon linewidth
$\Gamma_q$ is shown. The onset of a finite $\Gamma_q$ signals the
entrance of the spin-wave dispersion into the BF continuum and depending on the momentum
at which the entrance occurs the increase of $\Gamma_q$ is either smooth
(entrance at low momentum) or steep (entrance at high momentum). In
addition, we observe that $\Gamma_q$ has a maximum at the boundary of
the Brillouin zone for stronger interactions ($y=-0.75$ and $y=-1$ in
Fig.~\ref{BandW} respectively), whereas for smaller interactions we
observe the maximum at smaller momenta followed by a decrease of the
FWHM towards the zone boundary ($y=-2$ in Fig.~\ref{BandW}).

Interestingly, the coupling to the orbital degrees of freedom
does not only give rise to a featureless broadening of $S(q,\omega)$
but produces additional structures. These additional structures are
most obvious in Fig.~\ref{BFCymp75}(a). From
Fig.~\ref{BFCymp75}(b) it becomes clear that these structures are
dominated by local extrema as well as edges in the imaginary part of
the self-energy.  Eq.~(\ref{Imsig}) shows that such extrema can occur
if $\Omega^+_q(k_1,k_2)$, Eq.~(\ref{Omega_q}), becomes stationary as a
function of the momenta $k_1,k_2$ as long as the Heaviside functions
in Eq. (\ref{Imsig}) for these momenta are non-zero. The position of
the local maxima in the imaginary part of the self-energy is therefore
approximately given by the values $\Omega^+_q(k_1,k_2)$ at these
stationary points. These values correspond to the energy of
spin-orbital excitations into which the initial spin wave can decay, 
see Fig.~\ref{Feyn}(a) and which are stable against 
small redistributions of momenta. 

\begin{figure}[t!]
	  \includegraphics*[angle=-0,width=\linewidth]{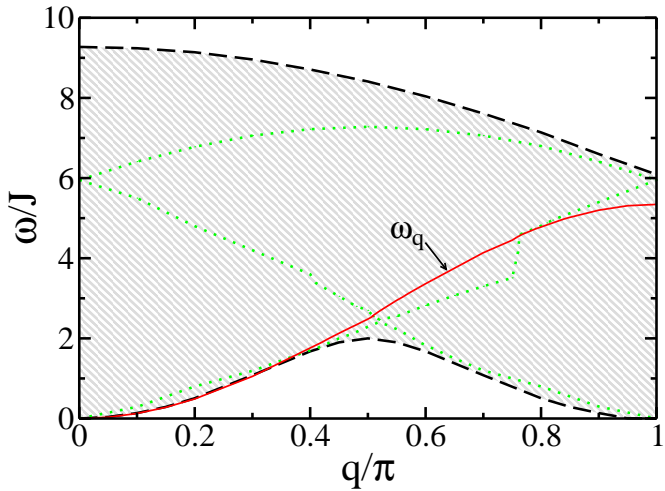}
\caption{(Color online) Effective dispersion of the spin wave $\omega_q$ for $y=-1$ 
(red solid line) together with coupled spin-orbital excitations 
deduced from $-\text{Im}\,\Sigma_2(q,\omega)$,
shown by (green) dots within the Bose-Fermi continuum (shadded area).}
\label{Disp}
\end{figure}

We conclude that while we do not have completely sharp spin-orbital
excitations any more as in the Hamiltonian (\ref{Dirac}) with FM
exchange considered in the introduction, there are still
characteristic spin-orbital excitations of finite width within the
spin-orbital continuum.  As shown in Fig.~\ref{Disp} we can extract
the dispersion of these characteristic excitations and find that the
coupled spin-orbital excitations are gapless for the parameters
considered here. However, as can be clearly seen in Figs.
\ref{BFCymp75}(b) and \ref{BFCymp75}(d), the weights of the low-energy
excitations are orders of magnitude smaller than the excitations
located at higher energies.  Hence the excitations at high energies
give the most dominant contribution to the dynamical spin structure
factor. We therefore expect that these excitations will generate
additional entropy in the corresponding temperature range which should
show up, for example, in the specific heat which will be studied in
the next section.

Moreover, we find that these coupled spin-orbtial excitations are
responsible for the Kohn anomalies mentioned above. We observe that
the Kohn anomalies at intermediate momenta occur when the energy of
the spin wave coincides with that of a characteristic spin-orbital
excitation. This is different from the Kohn anomaly in the spin-wave
dispersion of itinerant ferromagnets. In this case the interaction
between the localized spins given by the lattice ions is induced by
scattering with conduction electrons and hence the Kohn anomaly is
determined by the shape of the Fermi
surface.\cite{Kohn,Woll,Barnea,PhysRevLett.16.737,Halilov,Pajda,Moran}
The Kohn anomaly we find within the present context is also due to
interaction effects, where the nature of the interactions - coupled
spin-orbital degrees of freedom - is distinct from the ones of the
itinerant ferromagnets. For the case $x=-y=1$ the spin-wave dispersion
has a discontinuity of the order $\Delta\omega\simeq 0.01J$ at the
point $q\simeq 0.509\pi$ (see inset of Fig.~\ref{RSW}). However, for
the crossing points of the magnon and the coupled spin-orbital
excitation located at $q\simeq0.37\pi$ and $q\simeq0.76\pi$ (see
Fig.~\ref{Disp}) no Kohn-anomaly could be resolved. We believe that
this is a consequence of the weight of the coupled spin-orbital
excitations: Whereas at $q\simeq0.509\pi$ the imaginary part of the
self-energy displays a steep increase of several magnitudes, at the
other crossing points the slope towards the local maxima is far more
moderate.

\begin{figure}[t!]
\includegraphics*[angle=-0,width=\columnwidth]{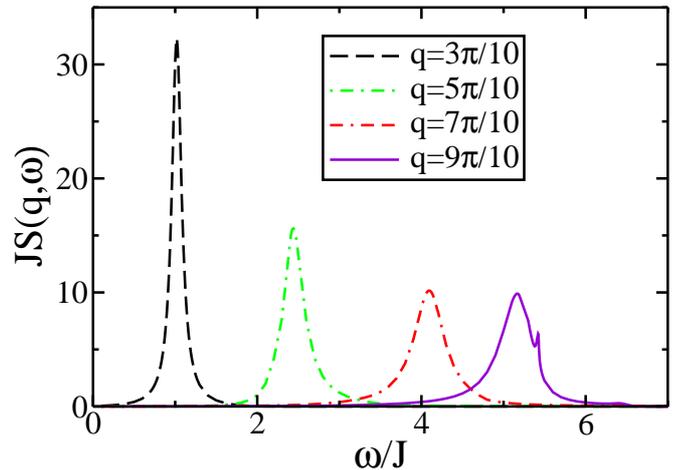}
\caption{(Color online) Dynamical spin structure factor $S(q,\omega)$ calculated 
perturbatively for the spin-orbital chain at temperature $T/J=0.1$ with $y=-1$.}
\label{DSFTym1}
\end{figure}

At finite temperatures two effects contribute to the broadening of
the central peaks of the dynamical spin structure factor. First, there is a broadening due to thermally excited
magnons which is already present in the 1D Heisenberg chain
discussed in Sec.~\ref{Sqw_uni}.  This is combined with the
broadening due to the interaction with the orbital degrees of freedom.
Here the BF continuum is smeared out by thermal fluctuations compared to the
zero temperature case. Results for the structure factor at finite
temperatures are shown in Fig.~\ref{DSFTym1}.

The broadening at small momenta is dominated by thermal fluctuations
and the lineshape is very similar to that of the pure spin model
discussed in Sec.~\ref{Sqw_uni}. A further strong broadening
in going from $q=0.3\pi$ to $q=0.5\pi$ signals the relevance of
coupled spin-orbital degrees of freedom on the spin dynamics at
intermediate and high momenta. Again, additional structures in
$S(q,\omega)$ are visible related to the spin-orbital excitations
discussed above. 

\begin{figure}[t!]
		\includegraphics*[angle=-90,width=\columnwidth]{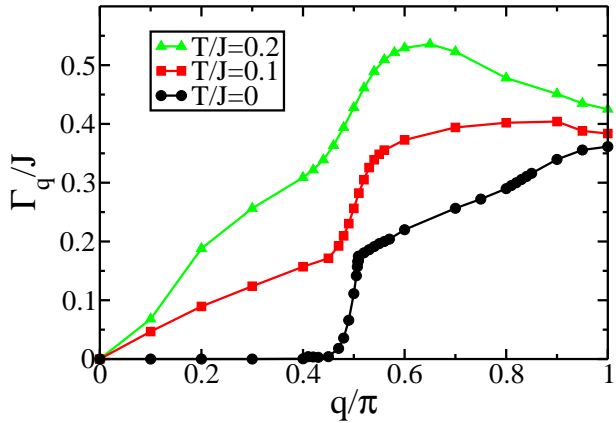}
\caption{(Color online) Magnon linewidth $\Gamma_q$ (FWHM of $S(q,\omega)$)  as 
a function of $q$, as obtained at $y=-1$ and temperatures $T/J=0$, 0.1 
and 0.2. The lines are guides to the eye.}
\label{BWT}
\end{figure}

Finally, we analyze the variation of the FWHM with increasing
temperature for a representative value of $y=-1$, see
Fig.~\ref{BWT}. 
At $T>0$ the thermal broadening at small momenta is clearly visible.
As in the zero temperature case another strong increase of $\Gamma_q$
between $q=2\pi/5$ and $q=\pi/2$ is observed due to coupled
spin-orbital degrees of freedom. For $T/J=0.1$ and $T/J=0.2$ we
observe that $\Gamma_q$ has a temperature dependent maximum from where
$\Gamma_q$ decreases towards the boundary of the Brillouin zone. This
is due to the fact that the thermal broadening of the central peaks of
the dynamic spin structure factor decreases from intermediate to high
momenta (see Fig.~\ref{-2ImGo}). Actually without coupling to any
orbital degrees of freedom we expect $\Gamma_q$ to be very small at
the boundary of the Brillouin zone. Thus a large bandwidth at $q=\pi$
makes the spin-orbital model distinct from a pure 1D Heisenberg
ferromagnet.

\section{Thermodynamics}
\label{C_PT}

Coupled spin-orbital degrees of freedom will not only influence
the spin dynamics but also the thermodynamics of the system. We
expect, in particular, that the spin-orbital excitations which were
shown to affect the dynamical spin structure factor in the previous section will
also become observable in thermodynamic quantities when comparing the
MF decoupling and the perturbative solution. In order to investigate
this issue we rewrite Eq.~(\ref{HGamma}) as
\begin{equation}\label{HT}
  \mathcal{H}_{S\tau}(1)=H_\textnormal{MF}+\delta H,
\end{equation}
with $\delta H=\mathcal{H}_{S\tau}(1)-H_\textnormal{MF}$. The MF part 
reads
\begin{equation}
\begin{aligned}
H_\textnormal{MF}=\sum_{j=1}^N&\left\{\mathcal{J}_\tau\left[
\vec{\tau}_j\cdot\vec{\tau}_{j+1}\right]_1\right.\left.
+\mathcal{J}_S\vec{S}_j\cdot\vec{S}_{j+1}\right.\\
&\left.-\left<\vec{S}_j\cdot\vec{S}_{j+1}\right>_{\textnormal{MF}}
\left<\left[\vec{\tau}_j\cdot\vec{\tau}_{j+1}\right]_1
\right>_{\textnormal{MF}}\right\}.
\end{aligned}
\end{equation}
The exchange constants $\mathcal{J}_{S,\tau}$ are defined in Eqs.
(\ref{Selfcons}).\cite{Remark} We use the
Hamiltonian (\ref{HT}) to determine the free energy per site
perturbatively, following the expansion,
\begin{equation}\label{FE}
f=f^S_\textnormal{MF}+f^\tau_\textnormal{MF}+\frac{1}{N}
\left<\delta H\right>_{\textnormal{MF}}^{c}\\
-\frac{1}{2NT}\left<\delta H^2\right>_{\textnormal{MF}}^{c}+\dots,
\end{equation}
where $f_\text{MF}^S$ ($f_\text{MF}^\tau$) is the expression for the
free energy per site stemming from the spin (pseudospin) sector within
the MF decoupling solution. The subscript indicates that the
respective correlation functions are calculated with $H_\text{MF}$. 
Moreover the
superscript $c$ means that the above expansion of the free energy is 
restricted to connected diagrams. We note that Eq.~(\ref{FE}) is a high 
temperature expansion valid if $|x|T/J\gg 1$ and $|y|T/J\gg 1$. 

A straightforward calculation shows that the first order contribution 
only shifts the free energy, Eq.~(\ref{FE}), and will not show up in
thermodynamic observables obtained by taking derivatives of the free
energy. For the second order contribution we have to evaluate two- and
four-point correlation functions both for the spin and the pseudospin part. 
\begin{figure}[t!]
\includegraphics{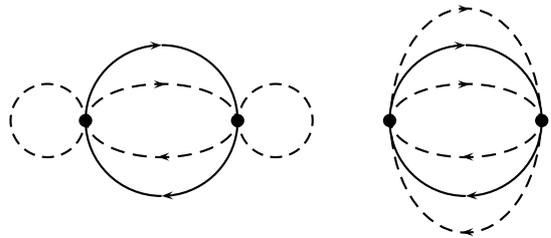}
\caption{Diagramatic representations of the second order
contributions to the free energy as given by Eq.~(\ref{2ndord}).}
\label{secordT}
\end{figure}
We use the abbreviations
\begin{equation}\label{SSSS}
\left<\left(\vec{S}_j\cdot\vec{S}_{j+1}\right)
\left(\vec{S}_l\cdot\vec{S}_{l+1}\right)\right>=a+b(j,l)
\end{equation}
for the spins and
\begin{equation}
\left<\left[\vec{\tau}_j\cdot\vec{\tau}_{j+1}\right]_1
\left[\vec{\tau}_l\cdot\vec{\tau}_{l+1}\right]_1\right>=c+d(j,l)
\end{equation}
for the pseudospins. Here the site-independent quantities $a$ and $c$ 
stand for the disconnected parts of
the four-point correlation functions whereas $b(j,l)$ and $d(j,l)$ follow 
from the connected ones. One finds after a
straightforward calculation that only the product of the
connected parts contributes to the second order correction, leading to
\begin{equation}\label{2ndord}
	\left<\delta H^2\right>_{\text{MF}}=J^2\sum_{j,l=1}^Nb(j,l)d(j,l)\,.
\end{equation}
To proceed further, we again apply the MSWT to the spin and a Jordan-Wigner transformation to the
pseudospin part. The evaluation of $d(j,l)$ is again straightforward and
yields
\begin{eqnarray}\label{djl}
d(j,l)&=&\frac{1}{2N^2}\sum_{k_1,k_2}n_\text{F}[\omega^\text{MF}_\text{F}(k_1,0)]
\{1-n_\text{F}[\omega^\text{MF}_\text{F}(k_2,0)]\}\nonumber \\
&&\hskip .7cm\times\textnormal{e}^{i(k_1-k_2)(j-l)}\left\{1+\cos(k_1+k_2)\right\}\nonumber\\
\end{eqnarray}
with $\omega^\text{MF}_\text{F}(q,\delta)$ as given in Eq.~(\ref{disfermMF}).
\begin{figure}[t!]
	\includegraphics*[width=1.0\columnwidth]{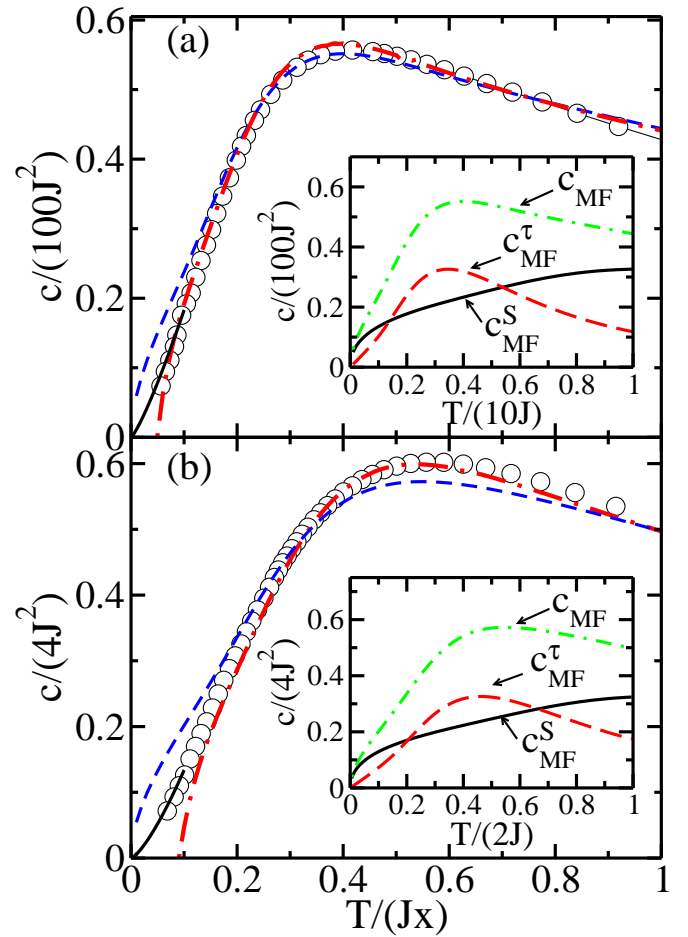}
        \caption{Specific heat per site $c/(Jx)^2$ as a function of
          $T/Jx$ for (a) $x=10$ and (b) $x=2$. The circles denote the
          numerical data from TMRG with the solid line obtained
          by a low-temperature fit of the TMRG data for the inner
          energy. The dashed lines correspond to MF decoupling while
          the dashed-dotted lines are the results obtained by perturbation
          theory.  The perturbative results are expected to be valid
          for $1/x^2\ll T/(Jx)\ll 1$.  Insets: Specific heat $c_\text{MF}$ within
          the MF solution (dashed-dotted line) with the contributions
          from the spin ($c_\text{MF}^S$ solid line) and the orbital ($c_\text{MF}^\tau$ dashed line)
          sector shown separately.}
\label{SH}
\end{figure}

The evaluation of Eq.~(\ref{SSSS}) is more involved. We first apply a
Dyson-Maleev transformation and treat the obtained expressions using
Wick's theorem. In addition, we also have to account for the constraint
of nonzero magnetization at $T>0$ imposed by the MSWT. 
The corresponding diagrams are shown in
Fig.~\ref{secordT}.
Within this approximation the specific heat per site reads
\begin{equation}
\label{cv}
c=c_\text{MF}+c_2+\dots\,,
\end{equation}
with
\begin{equation}\label{2ndcorr}
c_2=J^2T\frac{\partial^2}{\partial T^2}
\frac{\sum\limits_{j,l=1}^Nb(j,l)d(j,l)}{2NT}\,.
\end{equation}

\begin{figure}[t!]
	\includegraphics*[width=1.0\columnwidth]{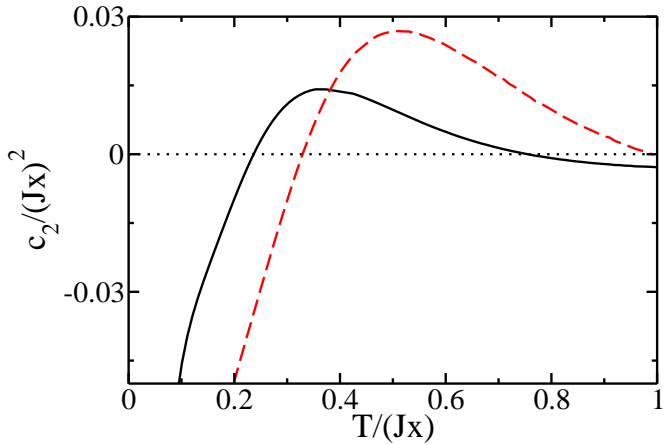}
	\caption{(Color online) Perturbative contribution to the specific heat per site $c_2/(Jx)^2$, Eq.~(\ref{2ndcorr}), as a function of $T/Jx$ for $x=10$
          (solid line) and $x=2$ (dashed line).}\label{contribs}
\end{figure}
We calculate the first term in Eq.~(\ref{cv}) 
within the MF decoupling, $c_\text{MF}=c_\text{MF}^S+c_\text{MF}^\tau$,
from the internal energy which is determined by the respective nearest-neighbor correlation functions
allowing us to keep terms up to quartic order in the bosonic
operators.\cite{ICM} This strategy makes it possible to obtain
reliable results for $c_\textnormal{MF}^S$ up to
$T/(|J_S|S^2)\leq1$. The second order correction $c_2$ given in 
Eq.~(\ref{2ndcorr}) is obtained
using the Dyson-Maleev transformation so that quartic terms are also
included and the order of approximation is the same. Since we are
using a high temperature expansion, Eq.~(\ref{FE}), in combination with the
MSWT to evaluate the diagrams, our results are only valid in an
intermediate temperature regime. If we restrict ourselves to
parameters $x=-y>0$ then this temperature range is given by $1/x\ll T/J
\ll xS^2$. In the following we therefore only consider the case $x\gg
1$ and compare the results from perturbation theory with numerical data
obtained by TMRG.

As shown in Fig.~\ref{SH}, the specific heat $c/(Jx)^2$ exhibits a
broad maximum which corresponds to the characteristic energies of spin
and fermionic particle-hole excitations. In the temperature range
where the perturbative approach is valid we find excellent agreement
with the numerical solution. In particular, the perturbative
correction $c_2$ (see Fig.~\ref{contribs}) correctly captures the
weight shift from low to intermediate temperatures visible when
comparing the numerical and the MF decoupling solution.  In spite of
this weight shift, the MF decoupling yields overall a very reasonable
description of the specific heat for both cases shown in
Fig.~\ref{SH}. For $x=10$, (Fig.~\ref{SH}(a)) the specific heat $c$ has a broad maximum at $T/(Jx)\simeq0.4$. This maximum results from a distinct maximum in the orbital contribution $c_\text{MF}^\tau$, see insets in Fig.~\ref{SH}. In contrast the spin contribution $c_\text{MF}^S$ increases steadily with increasing temperature, in agreement with the higher energy scale for spin excitations. As a result, the total specific heat has only a weaker and broader maximum than suggested by the orbital part.

The MF decoupling does seem to fail, however, at very low temperatures. Here the MF
solution predicts that spin excitations give the dominant contribution
leading to a $c(T)\sim\sqrt T$ behavior. This is in contrast to an
extrapolation of the numerical data shown as dot-dashed lines in
Fig.~\ref{SH} which suggests an approximately linear dependence on
temperature. This discrepancy comes as a surprise because our
perturbative calculations of the dynamical spin-structure factor in
Sec.~\ref{DSSF} lead us to the conclusion that the magnons survive as
sharp quasiparticles at low energies. More generally, one might argue
that the ground state does not show any entanglement between the two
sectors because of the classical nature of the FM state
thus allowing for spin-wave excitations at low energies. From the
point of view of a low-energy effective field theory, however, the
situation is much less clear. While a FM chain is described
by a low-energy effective theory with dynamical critical exponent
$z=2$, the fermionic orbital chain has $z=1$. A coupling of spatial
spin deviations to time-dependent orbital fluctuations then seems to
require that the low-energy effective theory for the coupled system
has $z=1$. As a consequence the temperature dependence of $c(T)$ would
indeed be linear. However, such an approach leaves open the role and
treatment of the Berry phase terms.

Within the perturbative approach the open question is whether or not
higher order contributions to the self-energy might induce a
significant broadening of the dynamical spin-structure factor also at
low energies. In this regard we note that the vertex responsible for
the broadening of $S(q,\omega)$ studied in Sec.~\ref{DSSF} does not
play any role for the thermodynamics of the system. Here the
constraint of vanishing magnetization means that such diagrams do not
contribute to {\it static} correlation functions so that the lowest
order corrections are caused by the vertex shown in
Fig.~\ref{secordT}.

\section{Conclusions}
\label{Conc}

In summary, we have investigated coupled spin-orbital degrees of
freedom in a one-dimensional model. For ferromagnetic exchange we have
shown that the considered model at special points in parameter space
can be written in terms of Dirac exchange operators for spin $S$ and
pseudospin $\tau$.  As a consequence, three collective excitations of
spin, orbital and coupled spin-orbital type do exist.  In particular, we discussed the case of Dirac exchange operators for $S=\tau=1/2$
where the dispersions of all three elementary excitations are
degenerate and lie within the spin-orbital continuum. While the
spin-orbital excitations stay confined in this case, a decay becomes
possible once we move away from this special point.

For antiferromagnetic exchange the one-dimensional spin-orbital model
captures fundamental aspects of physics relevant for transition metal
oxides and, as we have shown, sharp excitations do not exist. To
address the question how the spin dynamics is influenced by
fluctuating orbitals we considered the extreme quantum limit of
orbitals interacting via an XY-type coupling. This allowed us to map
the orbital sector onto free fermions using the Jordan-Wigner
transformation. In spin-wave theory the spin sector is described by
bosons so that our model corresponds to an
effective boson-fermion model which applies for low temperatures. An analytic calculation of the
properties within a mean-field decoupling approach is then
straightforward. Compared to a numerical phase diagram based on the
density-matrix renormalization group we find that the regime with
ferromagnetically polarized spins is stabilized by the decoupling
procedure. Furthermore, the mean-field decoupling gives rise to a
finite temperature dimerized phase for certain parameters when
starting from the ferromagnetic ground state. While a phase with
long-range dimer order at finite temperatures is not possible in a
purely one-dimensional model, the mean-field approach also completely
ignores any kind of coupled spin-orbital excitations.

Thus we developed a self-consistent perturbative scheme to explore the
role played by spin-orbital coupling. In perturbation theory the
boson-fermion interaction does not produce any infrared divergencies
in the one-dimensional model due to the lack of nesting. This makes a
perturbative calculation of the spin structure factor $S(q,\omega)$
possible. At large momenta $q$, we find that $S(q,\omega)$ shows a
significant broadening due to scattering of magnons by orbital
excitations. For small momenta, on the other hand, no broadening in
this lowest order perturbative approach is observed because the magnon
cannot scatter on these excitations. The onset of the broadening
occurs at momenta where the magnon enters the boson-fermion spectrum.
This point, as well as details of the full width at half maximum is
determined by the strength of interaction.  Most interestingly,
$S(q,\omega)$ does show additional peaks and shoulders corresponding
to characteristic spin-orbital excitations. At points where the
renormalized spin-wave dispersion and the dispersion of these
excitations cross, Kohn anomalies do occur.

Furthermore, we compared numerical data for the specific heat of the
spin-orbital model with the mean-field decoupling solution and an
approach where we also took the second order correction to the
mean-field result into account. Overall, we found that the mean-field
decoupling does describe the specific heat reasonably well. A
redistribution of entropic weight from low to intermediate
temperatures observed when comparing the numerical data and the
mean-field solution is very well captured by the second order
perturbative correction. An interesting open point is the behavior of
the specific heat $c(T)$ at low temperatures. While the mean-field
solution predicts $c(T)\sim\sqrt T$ due to spin-wave excitations, the
numerical data suggest instead that $c(T)\sim T$. We have speculated that a
coupling of the two sectors might indeed lead to a low-energy
effective theory with dynamical critical exponent $z=1$ but details of
such a theory need to be worked out in the future.

In conclusion, we have shown that while collective spin-orbital
excitations with infinite lifetime do not exist for antiferromagnetic
superexchange the coupled spin-orbital degrees of freedom have a strong influence
on the spin excitation spectrum as well as on the thermodynamic
properties of the system. However, the treatment of spin-orbital systems beyond the range of validity of mean-field decoupling and perturbative schemes remains an open problem in theory.

\acknowledgments

The authors thank O. P. Sushkov for valuable discussions. A.M.O. acknowledges support by the Foundation for Polish Science
(FNP) and by the Polish Ministry of Science and Higher Education
under Project No. N202 069639. J.S. acknowledges support by the graduate school of excellence MAINZ (MATCOR).

\appendix*
\section{Details of the perturbative approach for the boson-fermion
  model}

We start by a MF decoupling, rewriting the interaction as
$$
	H_1=H_{1,\text{MF}}+\underbrace{(H_1-H_{1,\text{MF}})}_{\delta H},
$$
with
$$
	\begin{aligned}
		H_{1,\text{MF}}=&\frac{1}{N^2}\sum_{k}\omega_\text{BF}(k,k,q)^2\left[\tilde n_{\text{b},k}f_q^\dagger f_q^{\ }+\tilde n_{\text{f},q}b_k^\dagger b_k^{\ }\right].
	\end{aligned}
$$
We treat $\delta H$ as perturbation. The averages $\tilde
n_{\text{a},p}=\bigl<a_p^\dagger a_p^{\ }\bigr>$ are determined
self-consistently within this MF scheme. This leads to a
renormalization of the magnon and fermion dispersions.  We find
\begin{equation}\label{remag}
	\zeta(q)=2JS\left(|y|-\frac{1}{N}\sum_k\cos k\ \tilde n_{\text{f},k}\right)(1-\cos q)-\mu,
\end{equation}
and
\begin{equation}\label{refer}
	\omega_\text{F}(q)=\left(S^2+x+\frac{2}{N}\sum_k(1-\cos k)\tilde n_{\text{b},k}\right)\cos q.
\end{equation}
For $T=0$ only the magnon dispersion is renormalized to
$$\zeta(q)=2JS(|y|+1/\pi)(1-\cos q).$$ By virtue of Dyson's equation
we may calculate the bosonic Green's function at $T=0$.\cite{FeWa}
Since the MF decoupling already takes the first order contributions
into account, the lowest order diagrams we obtain are of second order.
The self energy is thus approximated by the proper self-energy
obtained by summing up those diagrams which may be composed by the
second order diagrams. From this we have
$\Sigma(q,\omega)\approx\Sigma_2(q,\omega)\equiv
\Sigma_\text{BF}^+(q,\omega)+\Sigma_\text{BF}^-(q,\omega)+\Sigma_{2,1}(q)$
where the diagrams are given by $\Sigma_\text{BF}^+(q,\omega)$
(Fig.~\ref{Feyn}(a)), $\Sigma_\text{BF}^-(q,\omega)$
(Fig.~\ref{Feyn}(b)), and $\Sigma_{2,1}(q)$ (Fig.~\ref{Feyn}(c)). A
straigthforward calculation reveals that at zero temperature
$\Sigma_{2,1}(q)$ and $\Sigma_\text{BF}^-(q,\omega)$ vanish. For
$\Sigma_\text{BF}^+(q,\omega)$ we find the expression given in
Eq.~(\ref{sigma22}).

For finite temperatures we calculate the imaginary time Green's
function. We find 
\begin{equation}
\begin{aligned}
\Sigma_\text{BF}^\pm (q,\omega_{\nu,\textnormal{B}})&=-\frac{T^2}{N^2}
\sum_{k_1,k_2}\omega_\text{BF}^2(q,k_1,k_2)\\
&\times\sum_{a,b}\mathcal{G}^{(0)}_\textnormal{F}(k_2,\omega_{a,\textnormal{F}})
\mathcal{G}_\textnormal{F}^{(0)}(q-k_1+k_2,\omega_{b,\textnormal{F}})\\
&\times\mathcal{G}^{(0)}_\textnormal{B}(k_1,\pm(\omega_{\nu,\textnormal{B}}
-\omega_{b,\textnormal{F}}+\omega_{a,\textnormal{F}}))
\end{aligned}
\end{equation}
for the diagrams in Fig.~\ref{Feyn}(a,b),
and
\begin{equation}
\begin{aligned}
\Sigma_{2,1}(q)=&-\frac{T^2}{N^2}\sum_{k_1,k_2}\omega_\text{BF}(q,q,k_1)
\omega_\text{BF}(k_2,k_2,k_1)\\	
&\times\sum_{a,b}\left[\mathcal{G}_\textnormal{F}^{(0)}(k_1,\omega_{a,\textnormal{F}})
\right]^2\mathcal{G}_\textnormal{B}^{(0)}(k_2,\omega_{b,\textnormal{B}})
\end{aligned}
\end{equation}
for the diagram given in Fig.~\ref{Feyn}(c).
 Here we have used
$\mathcal{G}_\textnormal{F}^{(0)}(q,\omega_{\mu,\textnormal{F}})=
(i\omega_{\mu,\textnormal{F}}-\omega_F(q))^{-1}$ and $\mathcal{G}_\textnormal{B}^{(0)}(q,\omega_{\nu,\textnormal{B}})=
(i\omega_{\nu,\textnormal{B}}-\zeta(q))^{-1}$
as the fermionic and bosonic Matsubara Green's function for the noninteracting
Hamiltonian, respectively. $\omega_{\nu,\textnormal{F}}$ are the fermionic
Matsubara frequencies. After performing the frequency sums we end up
with
\begin{equation}
\begin{aligned}
\Sigma_{2,1}(q)=&-\frac{1}{TN^2}\sum_{k_1,k_2}\omega_\text{BF}(q,q,k_1)
\omega_\text{BF}(k_2,k_2,k_1)\\
&\times n_\textnormal{B}[\zeta(k_2)](1-n_\textnormal{F}[\omega_\text{F}(k_1)])
n_\textnormal{F}[\omega_\text{F}(k_1)],
\end{aligned}
\end{equation}
and the self-energies $\Sigma_\text{BF}^\pm(q,\omega_{\nu,\text{B}})$
as given in Eqs.~(\ref{SigmaBF}) and (\ref{SigmaBFm}).


\begin{thebibliography}{68}
\expandafter\ifx\csname natexlab\endcsname\relax\def\natexlab#1{#1}\fi
\expandafter\ifx\csname bibnamefont\endcsname\relax
  \def\bibnamefont#1{#1}\fi
\expandafter\ifx\csname bibfnamefont\endcsname\relax
  \def\bibfnamefont#1{#1}\fi
\expandafter\ifx\csname citenamefont\endcsname\relax
  \def\citenamefont#1{#1}\fi
\expandafter\ifx\csname url\endcsname\relax
  \def\url#1{\texttt{#1}}\fi
\expandafter\ifx\csname urlprefix\endcsname\relax\def\urlprefix{URL }\fi
\providecommand{\bibinfo}[2]{#2}
\providecommand{\eprint}[2][]{\url{#2}}

\bibitem[{\citenamefont{Landau}(1933)}]{Polaron}
\bibinfo{author}{\bibfnamefont{L.~D.} \bibnamefont{Landau}},
  \bibinfo{journal}{Phys. Z. Sowjetunion} \textbf{\bibinfo{volume}{3}},
  \bibinfo{pages}{664} (\bibinfo{year}{1933}).

\bibitem[{\citenamefont{Fr\"ohlich}(1954)}]{Polaron2}
\bibinfo{author}{\bibfnamefont{H.}~\bibnamefont{Fr\"ohlich}},
  \bibinfo{journal}{Advances in Physics} \textbf{\bibinfo{volume}{3}},
  \bibinfo{pages}{325} (\bibinfo{year}{1954}).

\bibitem[{\citenamefont{Peierls}(1955)}]{Peierls}
\bibinfo{author}{\bibfnamefont{R.~E.} \bibnamefont{Peierls}},
  \emph{\bibinfo{title}{Quantum Theory of Solids}} (\bibinfo{publisher}{Oxford
  University Press, Oxford}, \bibinfo{year}{1955}).

\bibitem[{\citenamefont{Das}(2003)}]{PhysRevLett.90.170403}
\bibinfo{author}{\bibfnamefont{K.~K.} \bibnamefont{Das}},
  \bibinfo{journal}{Phys. Rev. Lett.} \textbf{\bibinfo{volume}{90}},
  \bibinfo{pages}{170403} (\bibinfo{year}{2003}).

\bibitem[{\citenamefont{Albus et~al.}(2003)\citenamefont{Albus, Illuminati, and
  Eisert}}]{PhysRevA.68.023606}
\bibinfo{author}{\bibfnamefont{A.}~\bibnamefont{Albus}},
  \bibinfo{author}{\bibfnamefont{F.}~\bibnamefont{Illuminati}},
  \bibnamefont{and} \bibinfo{author}{\bibfnamefont{J.}~\bibnamefont{Eisert}},
  \bibinfo{journal}{Phys. Rev. A} \textbf{\bibinfo{volume}{68}},
  \bibinfo{pages}{023606} (\bibinfo{year}{2003}).

\bibitem[{\citenamefont{Cazalilla and Ho}(2003)}]{PhysRevLett.91.150403}
\bibinfo{author}{\bibfnamefont{M.~A.} \bibnamefont{Cazalilla}}
  \bibnamefont{and} \bibinfo{author}{\bibfnamefont{A.~F.} \bibnamefont{Ho}},
  \bibinfo{journal}{Phys. Rev. Lett.} \textbf{\bibinfo{volume}{91}},
  \bibinfo{pages}{150403} (\bibinfo{year}{2003}).

\bibitem[{\citenamefont{Mathey et~al.}(2004)\citenamefont{Mathey, Wang,
  Hofstetter, Lukin, and Demler}}]{PhysRevLett.93.120404}
\bibinfo{author}{\bibfnamefont{L.}~\bibnamefont{Mathey}},
  \bibinfo{author}{\bibfnamefont{D.-W.} \bibnamefont{Wang}},
  \bibinfo{author}{\bibfnamefont{W.}~\bibnamefont{Hofstetter}},
  \bibinfo{author}{\bibfnamefont{M.~D.} \bibnamefont{Lukin}}, \bibnamefont{and}
  \bibinfo{author}{\bibfnamefont{E.}~\bibnamefont{Demler}},
  \bibinfo{journal}{Phys. Rev. Lett.} \textbf{\bibinfo{volume}{93}},
  \bibinfo{pages}{120404} (\bibinfo{year}{2004}).

\bibitem[{\citenamefont{Lewenstein et~al.}(2004)\citenamefont{Lewenstein,
  Santos, Baranov, and Fehrmann}}]{PhysRevLett.92.050401}
\bibinfo{author}{\bibfnamefont{M.}~\bibnamefont{Lewenstein}},
  \bibinfo{author}{\bibfnamefont{L.}~\bibnamefont{Santos}},
  \bibinfo{author}{\bibfnamefont{M.~A.} \bibnamefont{Baranov}},
  \bibnamefont{and} \bibinfo{author}{\bibfnamefont{H.}~\bibnamefont{Fehrmann}},
  \bibinfo{journal}{Phys. Rev. Lett.} \textbf{\bibinfo{volume}{92}},
  \bibinfo{pages}{050401} (\bibinfo{year}{2004}).

\bibitem[{\citenamefont{Kugel and Khomskii}(1980)}]{KugKhom}
\bibinfo{author}{\bibfnamefont{K.~I.} \bibnamefont{Kugel}} \bibnamefont{and}
  \bibinfo{author}{\bibfnamefont{D.~I.} \bibnamefont{Khomskii}},
  \bibinfo{journal}{Sov. Phys. JETP} \textbf{\bibinfo{volume}{52}},
  \bibinfo{pages}{501} (\bibinfo{year}{1980}).

\bibitem[{\citenamefont{Feiner and Ole\'s}(1999)}]{FeinerOles}
\bibinfo{author}{\bibfnamefont{L.~F.} \bibnamefont{Feiner}} \bibnamefont{and}
  \bibinfo{author}{\bibfnamefont{A.~M.} \bibnamefont{Ole\'s}},
  \bibinfo{journal}{Phys. Rev. B} \textbf{\bibinfo{volume}{59}},
  \bibinfo{pages}{3295} (\bibinfo{year}{1999}).

\bibitem[{\citenamefont{Tokura and Nagaosa}(2000)}]{TokNag}
\bibinfo{author}{\bibfnamefont{Y.}~\bibnamefont{Tokura}} \bibnamefont{and}
  \bibinfo{author}{\bibfnamefont{N.}~\bibnamefont{Nagaosa}},
  \bibinfo{journal}{Science} \textbf{\bibinfo{volume}{288}},
  \bibinfo{pages}{462} (\bibinfo{year}{2000}).

\bibitem[{\citenamefont{Tokura}(2000)}]{CMR}
  \emph{\bibinfo{title}{Colossal Magnetoresistive Oxides}},\bibinfo{editor}{\bibfnamefont{ edited by Y.}~\bibnamefont{Tokura}},
  (\bibinfo{publisher}{Gordon and Breach, Amsterdam}, \bibinfo{year}{2000}).

\bibitem[{\citenamefont{Kovaleva et~al.}(2010)\citenamefont{Kovaleva, Ole\'s,
  Balbashov, Maljuk, Argyriou, Khaliullin, and Keimer}}]{KovalevaOles}
\bibinfo{author}{\bibfnamefont{N.~N.} \bibnamefont{Kovaleva}},
  \bibinfo{author}{\bibfnamefont{A.~M.} \bibnamefont{Ole\'s}},
  \bibinfo{author}{\bibfnamefont{A.~M.} \bibnamefont{Balbashov}},
  \bibinfo{author}{\bibfnamefont{A.}~\bibnamefont{Maljuk}},
  \bibinfo{author}{\bibfnamefont{D.~N.} \bibnamefont{Argyriou}},
  \bibinfo{author}{\bibfnamefont{G.}~\bibnamefont{Khaliullin}},
  \bibnamefont{and} \bibinfo{author}{\bibfnamefont{B.}~\bibnamefont{Keimer}},
  \bibinfo{journal}{Phys. Rev. B} \textbf{\bibinfo{volume}{81}},
  \bibinfo{pages}{235130} (\bibinfo{year}{2010}).

\bibitem[{\citenamefont{Khaliullin and Maekawa}(2000)}]{PhysRevLett.85.3950}
\bibinfo{author}{\bibfnamefont{G.}~\bibnamefont{Khaliullin}} \bibnamefont{and}
  \bibinfo{author}{\bibfnamefont{S.}~\bibnamefont{Maekawa}},
  \bibinfo{journal}{Phys. Rev. Lett.} \textbf{\bibinfo{volume}{85}},
  \bibinfo{pages}{3950} (\bibinfo{year}{2000}).

\bibitem[{\citenamefont{Keimer et~al.}(2000)\citenamefont{Keimer, Casa, Ivanov,
  Lynn, Zimmermann, Hill, Gibbs, Taguchi, and Tokura}}]{PhysRevLett.85.3946}
\bibinfo{author}{\bibfnamefont{B.}~\bibnamefont{Keimer}},
  \bibinfo{author}{\bibfnamefont{D.}~\bibnamefont{Casa}},
  \bibinfo{author}{\bibfnamefont{A.}~\bibnamefont{Ivanov}},
  \bibinfo{author}{\bibfnamefont{J.~W.} \bibnamefont{Lynn}},
  \bibinfo{author}{\bibfnamefont{M.~v.} \bibnamefont{Zimmermann}},
  \bibinfo{author}{\bibfnamefont{J.~P.} \bibnamefont{Hill}},
  \bibinfo{author}{\bibfnamefont{D.}~\bibnamefont{Gibbs}},
  \bibinfo{author}{\bibfnamefont{Y.}~\bibnamefont{Taguchi}}, \bibnamefont{and}
  \bibinfo{author}{\bibfnamefont{Y.}~\bibnamefont{Tokura}},
  \bibinfo{journal}{Phys. Rev. Lett.} \textbf{\bibinfo{volume}{85}},
  \bibinfo{pages}{3946} (\bibinfo{year}{2000}).

\bibitem[{\citenamefont{Hemberger et~al.}(2003)\citenamefont{Hemberger, von
  Nidda, Fritsch, Deisenhofer, Lobina, Rudolf, Lunkenheimer, Lichtenberg,
  Loidl, Bruns et~al.}}]{PhysRevLett.91.066403}
\bibinfo{author}{\bibfnamefont{J.}~\bibnamefont{Hemberger}},
  \bibinfo{author}{\bibfnamefont{H.-A.~K.} \bibnamefont{von Nidda}},
  \bibinfo{author}{\bibfnamefont{V.}~\bibnamefont{Fritsch}},
  \bibinfo{author}{\bibfnamefont{J.}~\bibnamefont{Deisenhofer}},
  \bibinfo{author}{\bibfnamefont{S.}~\bibnamefont{Lobina}},
  \bibinfo{author}{\bibfnamefont{T.}~\bibnamefont{Rudolf}},
  \bibinfo{author}{\bibfnamefont{P.}~\bibnamefont{Lunkenheimer}},
  \bibinfo{author}{\bibfnamefont{F.}~\bibnamefont{Lichtenberg}},
  \bibinfo{author}{\bibfnamefont{A.}~\bibnamefont{Loidl}},
  \bibinfo{author}{\bibfnamefont{D.}~\bibnamefont{Bruns}}, \bibnamefont{and}
  \bibinfo{author}{\bibfnamefont{B.}~\bibnamefont{B\"uchner}},
  \bibinfo{journal}{Phys. Rev. Lett.}
  \textbf{\bibinfo{volume}{91}}, \bibinfo{pages}{066403}
  (\bibinfo{year}{2003}).

\bibitem[{\citenamefont{Cwik et~al.}(2003)\citenamefont{Cwik, Lorenz, Baier,
  M\"uller, Andr\'e, Bour\'ee, Lichtenberg, Freimuth, Schmitz,
  M\"uller-Hartmann et~al.}}]{PhysRevB.68.060401}
\bibinfo{author}{\bibfnamefont{M.}~\bibnamefont{Cwik}},
  \bibinfo{author}{\bibfnamefont{T.}~\bibnamefont{Lorenz}},
  \bibinfo{author}{\bibfnamefont{J.}~\bibnamefont{Baier}},
  \bibinfo{author}{\bibfnamefont{R.}~\bibnamefont{M\"uller}},
  \bibinfo{author}{\bibfnamefont{G.}~\bibnamefont{Andr\'e}},
  \bibinfo{author}{\bibfnamefont{F.}~\bibnamefont{Bour\'ee}},
  \bibinfo{author}{\bibfnamefont{F.}~\bibnamefont{Lichtenberg}},
  \bibinfo{author}{\bibfnamefont{A.}~\bibnamefont{Freimuth}},
  \bibinfo{author}{\bibfnamefont{R.}~\bibnamefont{Schmitz}},
  \bibinfo{author}{\bibfnamefont{E.}~\bibnamefont{M\"uller-Hartmann}}, \bibnamefont{and}
  \bibinfo{author}{\bibfnamefont{M.}~\bibnamefont{Braden}}, \bibinfo{journal}{Phys. Rev. B}
  \textbf{\bibinfo{volume}{68}}, \bibinfo{pages}{060401}
  (\bibinfo{year}{2003}).

\bibitem[{\citenamefont{Ren et~al.}(1998)\citenamefont{Ren, Palstra, Khomskii,
  Nugroho, Menovsky, and Sawatzky}}]{TIMR1}
\bibinfo{author}{\bibfnamefont{Y.}~\bibnamefont{Ren}},
  \bibinfo{author}{\bibfnamefont{T.~T.~M.} \bibnamefont{Palstra}},
  \bibinfo{author}{\bibfnamefont{D.~I.} \bibnamefont{Khomskii}},
  \bibinfo{author}{\bibfnamefont{A.~A.} \bibnamefont{Nugroho}},
  \bibinfo{author}{\bibfnamefont{A.~A.} \bibnamefont{Menovsky}},
  \bibnamefont{and} \bibinfo{author}{\bibfnamefont{G.~A.}
  \bibnamefont{Sawatzky}}, \bibinfo{journal}{Nature (London)}
  \textbf{\bibinfo{volume}{396}}, \bibinfo{pages}{441} (\bibinfo{year}{1998}).

\bibitem[{\citenamefont{Ren et~al.}(2000)\citenamefont{Ren, Palstra, Khomskii,
  Nugroho, Menovsky, and Sawatzky}}]{TIMR2}
\bibinfo{author}{\bibfnamefont{Y.}~\bibnamefont{Ren}},
  \bibinfo{author}{\bibfnamefont{T.~T.~M.} \bibnamefont{Palstra}},
  \bibinfo{author}{\bibfnamefont{D.~I.} \bibnamefont{Khomskii}},
  \bibinfo{author}{\bibfnamefont{A.~A.} \bibnamefont{Nugroho}},
  \bibinfo{author}{\bibfnamefont{A.~A.} \bibnamefont{Menovsky}},
  \bibnamefont{and} \bibinfo{author}{\bibfnamefont{G.~A.}
  \bibnamefont{Sawatzky}}, \bibinfo{journal}{Phys. Rev. B}
  \textbf{\bibinfo{volume}{62}}, \bibinfo{pages}{6577} (\bibinfo{year}{2000}).

\bibitem[{\citenamefont{Noguchi et~al.}(2000)\citenamefont{Noguchi, Nakazawa,
  Oka, Arima, Wakabayashi, Nakao, and Murakami}}]{PhysRevB.62.R9271}
\bibinfo{author}{\bibfnamefont{M.}~\bibnamefont{Noguchi}},
  \bibinfo{author}{\bibfnamefont{A.}~\bibnamefont{Nakazawa}},
  \bibinfo{author}{\bibfnamefont{S.}~\bibnamefont{Oka}},
  \bibinfo{author}{\bibfnamefont{T.}~\bibnamefont{Arima}},
  \bibinfo{author}{\bibfnamefont{Y.}~\bibnamefont{Wakabayashi}},
  \bibinfo{author}{\bibfnamefont{H.}~\bibnamefont{Nakao}}, \bibnamefont{and}
  \bibinfo{author}{\bibfnamefont{Y.}~\bibnamefont{Murakami}},
  \bibinfo{journal}{Phys. Rev. B} \textbf{\bibinfo{volume}{62}},
  \bibinfo{pages}{R9271} (\bibinfo{year}{2000}).

\bibitem[{\citenamefont{Khaliullin et~al.}(2001)\citenamefont{Khaliullin,
  Horsch, and Ole\ifmmode~\acute{s}\else \'{s}\fi{}}}]{PhysRevLett.86.3879}
\bibinfo{author}{\bibfnamefont{G.}~\bibnamefont{Khaliullin}},
  \bibinfo{author}{\bibfnamefont{P.}~\bibnamefont{Horsch}}, \bibnamefont{and}
  \bibinfo{author}{\bibfnamefont{A.~M.} \bibnamefont{Ole\ifmmode~\acute{s}\else
  \'{s}\fi{}}}, \bibinfo{journal}{Phys. Rev. Lett.}
  \textbf{\bibinfo{volume}{86}}, \bibinfo{pages}{3879} (\bibinfo{year}{2001}).

\bibitem[{\citenamefont{Ulrich et~al.}(2003)\citenamefont{Ulrich, Khaliullin,
  Sirker, Reehuis, Ohl, Miyasaka, Tokura, and Keimer}}]{PhysRevLett.91.257202}
\bibinfo{author}{\bibfnamefont{C.}~\bibnamefont{Ulrich}},
  \bibinfo{author}{\bibfnamefont{G.}~\bibnamefont{Khaliullin}},
  \bibinfo{author}{\bibfnamefont{J.}~\bibnamefont{Sirker}},
  \bibinfo{author}{\bibfnamefont{M.}~\bibnamefont{Reehuis}},
  \bibinfo{author}{\bibfnamefont{M.}~\bibnamefont{Ohl}},
  \bibinfo{author}{\bibfnamefont{S.}~\bibnamefont{Miyasaka}},
  \bibinfo{author}{\bibfnamefont{Y.}~\bibnamefont{Tokura}}, \bibnamefont{and}
  \bibinfo{author}{\bibfnamefont{B.}~\bibnamefont{Keimer}},
  \bibinfo{journal}{Phys. Rev. Lett.} \textbf{\bibinfo{volume}{91}},
  \bibinfo{pages}{257202} (\bibinfo{year}{2003}).

\bibitem[{\citenamefont{Horsch et~al.}(2003)\citenamefont{Horsch, Khaliullin,
  and Ole\ifmmode~\acute{s}\else \'{s}\fi{}}}]{PhysRevLett.91.257203}
\bibinfo{author}{\bibfnamefont{P.}~\bibnamefont{Horsch}},
  \bibinfo{author}{\bibfnamefont{G.}~\bibnamefont{Khaliullin}},
  \bibnamefont{and} \bibinfo{author}{\bibfnamefont{A.~M.}
  \bibnamefont{Ole\ifmmode~\acute{s}\else \'{s}\fi{}}}, \bibinfo{journal}{Phys.
  Rev. Lett.} \textbf{\bibinfo{volume}{91}}, \bibinfo{pages}{257203}
  (\bibinfo{year}{2003}).

\bibitem[{\citenamefont{Sirker and Khaliullin}(2003)}]{PhysRevB.67.100408}
\bibinfo{author}{\bibfnamefont{J.}~\bibnamefont{Sirker}} \bibnamefont{and}
  \bibinfo{author}{\bibfnamefont{G.}~\bibnamefont{Khaliullin}},
  \bibinfo{journal}{Phys. Rev. B} \textbf{\bibinfo{volume}{67}},
  \bibinfo{pages}{100408} (\bibinfo{year}{2003}).

\bibitem[{\citenamefont{Horsch et~al.}(2008)\citenamefont{Horsch, Khaliullin,
  Ole\ifmmode~\acute{s}\else \'{s}\fi{}, and Feiner}}]{PhysRevLett.100.167205}
\bibinfo{author}{\bibfnamefont{P.}~\bibnamefont{Horsch}},
  \bibinfo{author}{\bibfnamefont{G.}~\bibnamefont{Khaliullin}},
  \bibinfo{author}{\bibfnamefont{A.~M.} \bibnamefont{Ole\ifmmode~\acute{s}\else
  \'{s}\fi{}}}, \bibnamefont{and} \bibinfo{author}{\bibfnamefont{L.~F.}
  \bibnamefont{Feiner}}, \bibinfo{journal}{Phys. Rev. Lett.}
  \textbf{\bibinfo{volume}{100}}, \bibinfo{pages}{167205}
  (\bibinfo{year}{2008}).

\bibitem[{\citenamefont{Sirker et~al.}(2008)\citenamefont{Sirker, Herzog,
  Ole\'{s}, and Horsch}}]{SPPRL}
\bibinfo{author}{\bibfnamefont{J.}~\bibnamefont{Sirker}},
  \bibinfo{author}{\bibfnamefont{A.}~\bibnamefont{Herzog}},
  \bibinfo{author}{\bibfnamefont{A.~M.} \bibnamefont{Ole\'{s}}},
  \bibnamefont{and} \bibinfo{author}{\bibfnamefont{P.}~\bibnamefont{Horsch}},
  \bibinfo{journal}{Phys. Rev. Lett.} \textbf{\bibinfo{volume}{101}},
  \bibinfo{pages}{157204} (\bibinfo{year}{2008}).

\bibitem[{\citenamefont{Ole\ifmmode~\acute{s}\else \'{s}\fi{}
  et~al.}(2006)\citenamefont{Ole\ifmmode~\acute{s}\else \'{s}\fi{}, Horsch,
  Feiner, and Khaliullin}}]{OlesHorsch}
\bibinfo{author}{\bibfnamefont{A.~M.} \bibnamefont{Ole\ifmmode~\acute{s}\else
  \'{s}\fi{}}}, \bibinfo{author}{\bibfnamefont{P.}~\bibnamefont{Horsch}},
  \bibinfo{author}{\bibfnamefont{L.~F.} \bibnamefont{Feiner}},
  \bibnamefont{and}
  \bibinfo{author}{\bibfnamefont{G.}~\bibnamefont{Khaliullin}},
  \bibinfo{journal}{Phys. Rev. Lett.} \textbf{\bibinfo{volume}{96}},
  \bibinfo{pages}{147205} (\bibinfo{year}{2006}).

\bibitem[{\citenamefont{Itoi et~al.}(2000{\natexlab{a}})\citenamefont{Itoi,
  Qin, and Affleck}}]{PhysRevB.61.6747}
\bibinfo{author}{\bibfnamefont{C.}~\bibnamefont{Itoi}},
  \bibinfo{author}{\bibfnamefont{S.}~\bibnamefont{Qin}}, \bibnamefont{and}
  \bibinfo{author}{\bibfnamefont{I.}~\bibnamefont{Affleck}},
  \bibinfo{journal}{Phys. Rev. B} \textbf{\bibinfo{volume}{61}},
  \bibinfo{pages}{6747} (\bibinfo{year}{2000}{\natexlab{a}}).

\bibitem[{\citenamefont{Sirker}(2004)}]{PhysRevB.69.104428}
\bibinfo{author}{\bibfnamefont{J.}~\bibnamefont{Sirker}},
  \bibinfo{journal}{Phys. Rev. B} \textbf{\bibinfo{volume}{69}},
  \bibinfo{pages}{104428} (\bibinfo{year}{2004}).

\bibitem[{\citenamefont{van~den Brink et~al.}(1998)\citenamefont{van~den Brink,
  Stekelenburg, Khomskii, Sawatzky, and Kugel}}]{PhysRevB.58.10276}
\bibinfo{author}{\bibfnamefont{J.}~\bibnamefont{van~den Brink}},
  \bibinfo{author}{\bibfnamefont{W.}~\bibnamefont{Stekelenburg}},
  \bibinfo{author}{\bibfnamefont{D.~I.} \bibnamefont{Khomskii}},
  \bibinfo{author}{\bibfnamefont{G.~A.} \bibnamefont{Sawatzky}},
  \bibnamefont{and} \bibinfo{author}{\bibfnamefont{K.~I.} \bibnamefont{Kugel}},
  \bibinfo{journal}{Phys. Rev. B} \textbf{\bibinfo{volume}{58}},
  \bibinfo{pages}{10276} (\bibinfo{year}{1998}).

\bibitem[{\citenamefont{Li et~al.}(1998)\citenamefont{Li, Ma, Shi, and
  Zhang}}]{LiMa}
\bibinfo{author}{\bibfnamefont{Y.~Q.} \bibnamefont{Li}},
  \bibinfo{author}{\bibfnamefont{M.}~\bibnamefont{Ma}},
  \bibinfo{author}{\bibfnamefont{D.~N.} \bibnamefont{Shi}}, \bibnamefont{and}
  \bibinfo{author}{\bibfnamefont{F.~C.} \bibnamefont{Zhang}},
  \bibinfo{journal}{Phys Rev Lett} \textbf{\bibinfo{volume}{81}},
  \bibinfo{pages}{3527} (\bibinfo{year}{1998}).

\bibitem[{rem({\natexlab{a}})}]{remarkSU4}
\bibinfo{note}{This is reminiscent to the SU(4) model with AF exchange where
  all three elementary excitations are degenerate. However in the latter case
  the elementary excitations are of coupled spin-orbital type, whereas in the
  present case we find elementary spin, orbital and combined spin-orbital
  excitations.}

\bibitem[{vdB()}]{vdBe}
\bibinfo{note}{This, we believe, is a consequence of a factor of $2$ missed in
  the analysis of the elementary dispersions in this work.}

\bibitem[{\citenamefont{Dirac}(1958)}]{Dirac}
\bibinfo{author}{\bibfnamefont{P.~A.~M.} \bibnamefont{Dirac}},
  \emph{\bibinfo{title}{Principles of Quantum Mechanics}}
  (\bibinfo{publisher}{Oxford University Press, Oxford}, \bibinfo{year}{1958}).

\bibitem[{\citenamefont{Schr\"odinger}(1941)}]{Schroedinger}
\bibinfo{author}{\bibfnamefont{E.}~\bibnamefont{Schr\"odinger}},
  \bibinfo{journal}{Proceedings of the Royal Irish Academy Section A}
  \textbf{\bibinfo{volume}{47}}, \bibinfo{pages}{39} (\bibinfo{year}{1941}).

\bibitem[{\citenamefont{Brown}(1985)}]{PhysRevB.31.3118}
\bibinfo{author}{\bibfnamefont{H.~A.} \bibnamefont{Brown}},
  \bibinfo{journal}{Phys. Rev. B} \textbf{\bibinfo{volume}{31}},
  \bibinfo{pages}{3118} (\bibinfo{year}{1985}).

\bibitem[{\citenamefont{Feiner et~al.}(1998)\citenamefont{Feiner, Ole\'s, and
  Zaanen}}]{Fei98}
\bibinfo{author}{\bibfnamefont{L.~F.} \bibnamefont{Feiner}},
  \bibinfo{author}{\bibfnamefont{A.~M.} \bibnamefont{Ole\'s}},
  \bibnamefont{and} \bibinfo{author}{\bibfnamefont{J.}~\bibnamefont{Zaanen}},
  \bibinfo{journal}{J. Phys.: Condens. Matter} \textbf{\bibinfo{volume}{10}},
  \bibinfo{pages}{L555} (\bibinfo{year}{1998}).

\bibitem[{\citenamefont{Ole\'s et~al.}(2000)\citenamefont{Ole\'s, Feiner, and
  Zaanen}}]{Ole00}
\bibinfo{author}{\bibfnamefont{A.~M.} \bibnamefont{Ole\'s}},
  \bibinfo{author}{\bibfnamefont{L.~F.} \bibnamefont{Feiner}},
  \bibnamefont{and} \bibinfo{author}{\bibfnamefont{J.}~\bibnamefont{Zaanen}},
  \bibinfo{journal}{Phys. Rev. B} \textbf{\bibinfo{volume}{61}},
  \bibinfo{pages}{6257} (\bibinfo{year}{2000}).

\bibitem[{\citenamefont{Takahashi}(1990)}]{MSWT2}
\bibinfo{author}{\bibfnamefont{M.}~\bibnamefont{Takahashi}},
  \bibinfo{journal}{Phys. Rev. B} \textbf{\bibinfo{volume}{42}},
  \bibinfo{pages}{766} (\bibinfo{year}{1990}).

\bibitem[{\citenamefont{Takahashi}(1993)}]{MSWT3}
\bibinfo{author}{\bibfnamefont{M.}~\bibnamefont{Takahashi}},
  \bibinfo{journal}{Phys. Rev. B} \textbf{\bibinfo{volume}{47}},
  \bibinfo{pages}{8336} (\bibinfo{year}{1993}).

\bibitem[{\citenamefont{Takahashi}(1987)}]{MSWT}
\bibinfo{author}{\bibfnamefont{M.}~\bibnamefont{Takahashi}},
  \bibinfo{journal}{Phys. Rev. Lett} \textbf{\bibinfo{volume}{58}},
  \bibinfo{pages}{168} (\bibinfo{year}{1987}).

\bibitem[{\citenamefont{Takahashi}(1986)}]{MSWTlong}
\bibinfo{author}{\bibfnamefont{M.}~\bibnamefont{Takahashi}},
  \bibinfo{journal}{Prog. Theor. Phys. Suppl.} \textbf{\bibinfo{volume}{87}},
  \bibinfo{pages}{233} (\bibinfo{year}{1986}).

\bibitem[{\citenamefont{Mizokawa and Fujimori}(1996)}]{PhysRevB.54.5368}
\bibinfo{author}{\bibfnamefont{T.}~\bibnamefont{Mizokawa}} \bibnamefont{and}
  \bibinfo{author}{\bibfnamefont{A.}~\bibnamefont{Fujimori}},
  \bibinfo{journal}{Phys. Rev. B} \textbf{\bibinfo{volume}{54}},
  \bibinfo{pages}{5368} (\bibinfo{year}{1996}).

\bibitem[{\citenamefont{Sawada and Terakura}(1998)}]{PhysRevB.58.6831}
\bibinfo{author}{\bibfnamefont{H.}~\bibnamefont{Sawada}} \bibnamefont{and}
  \bibinfo{author}{\bibfnamefont{K.}~\bibnamefont{Terakura}},
  \bibinfo{journal}{Phys. Rev. B} \textbf{\bibinfo{volume}{58}},
  \bibinfo{pages}{6831} (\bibinfo{year}{1998}).

\bibitem[{\citenamefont{Mizokawa et~al.}(1999)\citenamefont{Mizokawa, Khomskii,
  and Sawatzky}}]{PhysRevB.60.7309}
\bibinfo{author}{\bibfnamefont{T.}~\bibnamefont{Mizokawa}},
  \bibinfo{author}{\bibfnamefont{D.~I.} \bibnamefont{Khomskii}},
  \bibnamefont{and} \bibinfo{author}{\bibfnamefont{G.~A.}
  \bibnamefont{Sawatzky}}, \bibinfo{journal}{Phys. Rev. B}
  \textbf{\bibinfo{volume}{60}}, \bibinfo{pages}{7309} (\bibinfo{year}{1999}).

\bibitem[{\citenamefont{Park et~al.}(2000)\citenamefont{Park, Tjeng, Tanaka,
  Allen, Chen, Metcalf, Honig, de~Groot, and Sawatzky}}]{PhysRevB.61.11506}
\bibinfo{author}{\bibfnamefont{J.-H.} \bibnamefont{Park}},
  \bibinfo{author}{\bibfnamefont{L.~H.} \bibnamefont{Tjeng}},
  \bibinfo{author}{\bibfnamefont{A.}~\bibnamefont{Tanaka}},
  \bibinfo{author}{\bibfnamefont{J.~W.} \bibnamefont{Allen}},
  \bibinfo{author}{\bibfnamefont{C.~T.} \bibnamefont{Chen}},
  \bibinfo{author}{\bibfnamefont{P.}~\bibnamefont{Metcalf}},
  \bibinfo{author}{\bibfnamefont{J.~M.} \bibnamefont{Honig}},
  \bibinfo{author}{\bibfnamefont{F.~M.~F.} \bibnamefont{de~Groot}},
  \bibnamefont{and} \bibinfo{author}{\bibfnamefont{G.~A.}
  \bibnamefont{Sawatzky}}, \bibinfo{journal}{Phys. Rev. B}
  \textbf{\bibinfo{volume}{61}}, \bibinfo{pages}{11506} (\bibinfo{year}{2000}).

\bibitem[{\citenamefont{Bursill et~al.}(1996)\citenamefont{Bursill, Xiang, and
  Gehring}}]{BursillXiang}
\bibinfo{author}{\bibfnamefont{R.~J.} \bibnamefont{Bursill}},
  \bibinfo{author}{\bibfnamefont{T.}~\bibnamefont{Xiang}}, \bibnamefont{and}
  \bibinfo{author}{\bibfnamefont{G.~A.} \bibnamefont{Gehring}},
  \bibinfo{journal}{J. Phys. Cond. Mat.} \textbf{\bibinfo{volume}{8}},
  \bibinfo{pages}{L583} (\bibinfo{year}{1996}).

\bibitem[{\citenamefont{Wang and Xiang}(1997)}]{WangXiang}
\bibinfo{author}{\bibfnamefont{X.}~\bibnamefont{Wang}} \bibnamefont{and}
  \bibinfo{author}{\bibfnamefont{T.}~\bibnamefont{Xiang}},
  \bibinfo{journal}{Phys Rev B} \textbf{\bibinfo{volume}{56}},
  \bibinfo{pages}{5061} (\bibinfo{year}{1997}).

\bibitem[{\citenamefont{Sirker and Kl\"umper}(2002)}]{SirkerKluemperEPL}
\bibinfo{author}{\bibfnamefont{J.}~\bibnamefont{Sirker}} \bibnamefont{and}
  \bibinfo{author}{\bibfnamefont{A.}~\bibnamefont{Kl\"umper}},
  \bibinfo{journal}{Europhys. Lett.} \textbf{\bibinfo{volume}{60}},
  \bibinfo{pages}{262} (\bibinfo{year}{2002}).

\bibitem[{\citenamefont{White and Huse}(1993)}]{WhiteHuse}
\bibinfo{author}{\bibfnamefont{S.~R.} \bibnamefont{White}} \bibnamefont{and}
  \bibinfo{author}{\bibfnamefont{D.~S.} \bibnamefont{Huse}},
  \bibinfo{journal}{Phys. Rev. B} \textbf{\bibinfo{volume}{48}},
  \bibinfo{pages}{3844} (\bibinfo{year}{1993}).

\bibitem[{\citenamefont{Miyashita and Kawakami}(2005)}]{Kawakami}
\bibinfo{author}{\bibfnamefont{S.}~\bibnamefont{Miyashita}} \bibnamefont{and}
  \bibinfo{author}{\bibfnamefont{N.}~\bibnamefont{Kawakami}},
  \bibinfo{journal}{J. Phys. Soc. Japan} \textbf{\bibinfo{volume}{74}},
  \bibinfo{pages}{758} (\bibinfo{year}{2005}).

\bibitem[{\citenamefont{Itoi et~al.}(2000{\natexlab{b}})\citenamefont{Itoi,
  Qin, and Affleck}}]{ItoiQin}
\bibinfo{author}{\bibfnamefont{C.}~\bibnamefont{Itoi}},
  \bibinfo{author}{\bibfnamefont{S.}~\bibnamefont{Qin}}, \bibnamefont{and}
  \bibinfo{author}{\bibfnamefont{I.}~\bibnamefont{Affleck}},
  \bibinfo{journal}{Phys. Rev. B} \textbf{\bibinfo{volume}{61}},
  \bibinfo{pages}{6747} (\bibinfo{year}{2000}{\natexlab{b}}).

\bibitem[{\citenamefont{Chen et~al.}(2007)\citenamefont{Chen, Wang, Li, and
  Zhang}}]{ChenWang}
\bibinfo{author}{\bibfnamefont{Y.}~\bibnamefont{Chen}},
  \bibinfo{author}{\bibfnamefont{Z.~D.} \bibnamefont{Wang}},
  \bibinfo{author}{\bibfnamefont{Y.~Q.} \bibnamefont{Li}}, \bibnamefont{and}
  \bibinfo{author}{\bibfnamefont{F.~C.} \bibnamefont{Zhang}},
  \bibinfo{journal}{Phys. Rev. B} \textbf{\bibinfo{volume}{75}},
  \bibinfo{pages}{195113} (\bibinfo{year}{2007}).

\bibitem[{\citenamefont{Herzog et~al.}(2010)\citenamefont{Herzog, Horsch,
  Ole\'{s}, and Sirker}}]{ICM}
\bibinfo{author}{\bibfnamefont{A.}~\bibnamefont{Herzog}},
  \bibinfo{author}{\bibfnamefont{P.}~\bibnamefont{Horsch}},
  \bibinfo{author}{\bibfnamefont{A.~M.} \bibnamefont{Ole\'{s}}},
  \bibnamefont{and} \bibinfo{author}{\bibfnamefont{J.}~\bibnamefont{Sirker}},
  \bibinfo{journal}{Journal of Physics Conference Series}
  \textbf{\bibinfo{volume}{200}}, \bibinfo{pages}{022017}
  (\bibinfo{year}{2010}).

\bibitem[{\citenamefont{Pincus}(1971)}]{Pincus19711971}
\bibinfo{author}{\bibfnamefont{P.}~\bibnamefont{Pincus}},
  \bibinfo{journal}{Solid State Communications} \textbf{\bibinfo{volume}{9}},
  \bibinfo{pages}{1971 } (\bibinfo{year}{1971}).

\bibitem[{rem({\natexlab{b}})}]{remarkdim}
\bibinfo{note}{The discussion of the dynamical spin structure factor for the
  dimerized FM chain will be given elsewhere.}

\bibitem[{FT0()}]{FT0}
\bibinfo{note}{We note that the MF decoupling applied to the interaction given
  in Eq.~(\ref{HC2}) does not renormalize the dispersion of the fermions at
  $T=0$.}

\bibitem[{\citenamefont{Giamarchi}(2003)}]{Giamarchi}
\bibinfo{author}{\bibfnamefont{T.}~\bibnamefont{Giamarchi}},
  \emph{\bibinfo{title}{Quantum Physics in One Dimension}}
  (\bibinfo{publisher}{Clarendon Press, Oxford}, \bibinfo{year}{2003}).

\bibitem[{\citenamefont{Metzner et~al.}(1998)\citenamefont{Metzner, Castellani,
  and Castro}}]{Metzner}
\bibinfo{author}{\bibfnamefont{W.}~\bibnamefont{Metzner}},
  \bibinfo{author}{\bibfnamefont{C.}~\bibnamefont{Castellani}},
  \bibnamefont{and} \bibinfo{author}{\bibfnamefont{C.~D.}
  \bibnamefont{Castro}}, \bibinfo{journal}{Advances in Physics}
  \textbf{\bibinfo{volume}{47}}, \bibinfo{pages}{317} (\bibinfo{year}{1998}).

\bibitem[{\citenamefont{Kohn}(1959)}]{Kohn}
\bibinfo{author}{\bibfnamefont{W.}~\bibnamefont{Kohn}}, \bibinfo{journal}{Phys.
  Rev. Lett.} \textbf{\bibinfo{volume}{2}}, \bibinfo{pages}{393}
  (\bibinfo{year}{1959}).

\bibitem[{\citenamefont{Woll and Nettel}(1961)}]{Woll}
\bibinfo{author}{\bibfnamefont{E.~J.} \bibnamefont{Woll}} \bibnamefont{and}
  \bibinfo{author}{\bibfnamefont{S.~J.} \bibnamefont{Nettel}},
  \bibinfo{journal}{Phys. Rev.} \textbf{\bibinfo{volume}{123}},
  \bibinfo{pages}{769} (\bibinfo{year}{1961}).

\bibitem[{\citenamefont{Barnea and Horwitz}(1975)}]{Barnea}
\bibinfo{author}{\bibfnamefont{G.}~\bibnamefont{Barnea}} \bibnamefont{and}
  \bibinfo{author}{\bibfnamefont{G.}~\bibnamefont{Horwitz}},
  \bibinfo{journal}{J. Phys. C} \textbf{\bibinfo{volume}{8}},
  \bibinfo{pages}{2124} (\bibinfo{year}{1975}).

\bibitem[{\citenamefont{M\o{}ller and Houmann}(1966)}]{PhysRevLett.16.737}
\bibinfo{author}{\bibfnamefont{H.~B.} \bibnamefont{M\o{}ller}}
  \bibnamefont{and} \bibinfo{author}{\bibfnamefont{J.~C.~G.}
  \bibnamefont{Houmann}}, \bibinfo{journal}{Phys. Rev. Lett.}
  \textbf{\bibinfo{volume}{16}}, \bibinfo{pages}{737} (\bibinfo{year}{1966}).

\bibitem[{\citenamefont{Halilov et~al.}(1998)\citenamefont{Halilov, Eschrig,
  Perlov, and Oppeneer}}]{Halilov}
\bibinfo{author}{\bibfnamefont{S.~V.} \bibnamefont{Halilov}},
  \bibinfo{author}{\bibfnamefont{H.}~\bibnamefont{Eschrig}},
  \bibinfo{author}{\bibfnamefont{A.~Y.} \bibnamefont{Perlov}},
  \bibnamefont{and} \bibinfo{author}{\bibfnamefont{P.~M.}
  \bibnamefont{Oppeneer}}, \bibinfo{journal}{Phys. Rev. B}
  \textbf{\bibinfo{volume}{58}}, \bibinfo{pages}{293} (\bibinfo{year}{1998}).

\bibitem[{\citenamefont{Pajda et~al.}(2001)\citenamefont{Pajda, Kudrnovsk\'y,
  Turek, Drchal, and Bruno}}]{Pajda}
\bibinfo{author}{\bibfnamefont{M.}~\bibnamefont{Pajda}},
  \bibinfo{author}{\bibfnamefont{J.}~\bibnamefont{Kudrnovsk\'y}},
  \bibinfo{author}{\bibfnamefont{I.}~\bibnamefont{Turek}},
  \bibinfo{author}{\bibfnamefont{V.}~\bibnamefont{Drchal}}, \bibnamefont{and}
  \bibinfo{author}{\bibfnamefont{P.}~\bibnamefont{Bruno}},
  \bibinfo{journal}{Phys. Rev. B} \textbf{\bibinfo{volume}{64}},
  \bibinfo{pages}{174402} (\bibinfo{year}{2001}).

\bibitem[{\citenamefont{Mor\'an et~al.}(2003)\citenamefont{Mor\'an, Ederer, and
  F\"ahnle}}]{Moran}
\bibinfo{author}{\bibfnamefont{S.}~\bibnamefont{Mor\'an}},
  \bibinfo{author}{\bibfnamefont{C.}~\bibnamefont{Ederer}}, \bibnamefont{and}
  \bibinfo{author}{\bibfnamefont{M.}~\bibnamefont{F\"ahnle}},
  \bibinfo{journal}{Phys. Rev. B} \textbf{\bibinfo{volume}{67}},
  \bibinfo{pages}{012407} (\bibinfo{year}{2003}).

\bibitem[{Rem()}]{Remark}
\bibinfo{note}{Performing the MF decoupling with the Hamiltonian (\ref{HGamma})
  also takes the quartic orders of the Dyson-Maleev transformation into
  account. These higher orders are neglected in the MF decoupling performed in
  Sec.~\ref{PT}. For low temperatures, i.e. the temperature region we have used
  in Sec.~\ref{PT} the contributions from these higher order terms are small.}

\bibitem[{\citenamefont{Fetter and Walecka}(1971)}]{FeWa}
\bibinfo{author}{\bibfnamefont{A.~L.} \bibnamefont{Fetter}} \bibnamefont{and}
  \bibinfo{author}{\bibfnamefont{J.~D.} \bibnamefont{Walecka}},
  \emph{\bibinfo{title}{Quantum Theory of Many-Particle Sytems}}
  (\bibinfo{publisher}{McGraw-Hill, Inc., New York}, \bibinfo{year}{1971}).

\end{thebibliography}

\end{document}